\documentclass[%
reprint,
superscriptaddress,
 amsmath,amssymb,
 aps,
]{revtex4-2}

\usepackage{graphicx}
\usepackage{dcolumn}
\usepackage{bm}
\usepackage{braket}
\usepackage{xcolor}
\usepackage{hyperref}
\hypersetup{
    colorlinks=true,
    linkcolor=blue,
    citecolor=blue,
    filecolor=blue,
    urlcolor=blue,
    pdftitle={Your Document Title},
    pdfauthor={Your Name},
    pdfsubject={},
    pdfkeywords={},
}

\begin{document}
%

\title{Efficient sampling using Macrocanonical Monte Carlo and density of states mapping}

\author{Jiewei Ding}
\affiliation{
Department of Physics, City University of Hong Kong, Kowloon, Hong Kong
}%
\author{Jiahao Su}
\affiliation{%
  School of Science, Harbin Institute of Technology, Shenzhen, 518055, China
}%
\author{Ho-Kin Tang}
\email{denghaojian@hit.edu.cn}
\affiliation{%
  School of Science, Harbin Institute of Technology, Shenzhen, 518055, China
}%

\author{Wing Chi Yu}
\email{wingcyu@cityu.edu.hk}
\affiliation{
Department of Physics, City University of Hong Kong, Kowloon, Hong Kong
}

\date{\today}

\begin{abstract}
In the context of Monte Carlo sampling for lattice models, the complexity of the energy landscape often leads to Markov chains being trapped in local optima, thereby increasing the correlation between samples and reducing sampling efficiency. This study proposes a Monte Carlo algorithm that effectively addresses the irregularities of the energy landscape through the introduction of the estimated density of states. This algorithm enhances the accuracy in the study of phase transitions and is not model-specific. Although our algorithm is primarily demonstrated on the two-dimensional square lattice model, the method is also applicable to a broader range of lattice and higher-dimensional models. Furthermore, the study develops a method for estimating the density of states of large systems based on that of smaller systems, enabling high-precision density of states estimation within specific energy intervals in large systems without sampling. For regions of lower precision, a re-weighting strategy is employed to adjust the density of states to enhance the precision further. This algorithm is not only significant within the field of lattice model sampling but may also inspire applications of the Monte Carlo method in other domains.
\end{abstract}

\maketitle


\section{Introduction}
In condensed matter physics, the Markov Chain Monte Carlo (MCMC) method plays a pivotal role, especially in exploring phase transitions and critical phenomena \cite{landau2021guide}. Improving the accuracy of phase transition exploration requires addressing the sources of error in MCMC, primarily stemming from the insufficient number of samples and autocorrelation \cite{landau2021guide}. One strategy to increase the number of samples involves accelerating parts of the simulation process using GPUs, though research in this area remains limited \cite{preis2009gpu,block2010multi,komura2012large}. In contrast, efforts to reduce autocorrelation have been more substantial. 
This is due to the challenge of inefficient sampling in complex models which exhibit severe autocorrelation. 
For instance, in the Potts model with first-order phase transitions \cite{wu1982potts}, significant energy barriers hinder efficient sampling across the entire energy space. To enhance the probability of crossing these barriers, methods such as Replica Exchange Monte Carlo and Multicanonical Monte Carlo have been proposed \cite{berg1992multicanonical,swendsen1986replica}. The former performs well in relatively smooth energy landscapes but requires an increased number of Markov chains for state exchanges in more complex systems. The latter is an effective method for overcoming energy barriers by recalibrating the weight of each state, thereby facilitating access to low-probability states. This approach significantly reduces the correlations between samples and biases in sampling outcomes. Its main limitation lies in the need to estimate the density of states and the potential for memory overflow during the recalibration. ~\\ 

Near the phase transition temperatures, many models encounter the phenomenon known as critical slowing down, where a decrease in the acceptance rate leads to a high correlation between successive spin configurations \cite{fehske2007computational}. To address this challenge, the Wolff algorithm, also known as the cluster update method, was introduced \cite{wolff1989collective}. However, this algorithm is primarily effective for cases that can be simplified to nearest-neighbor interacting models and are less suitable for models involving longer range of interactions or plaquette interactions \cite{liu2017self}. In strongly frustrated systems, such as the fully frustrated XY (FFXY) model \cite{teitel1983phase} and the Edwards-Anderson (EA) model \cite{nishimori2001statistical}, the Wolff algorithm is also limited in effectiveness \cite{cataudella1996efficient}. Recently, a novel deep learning approach known as the self-learning cluster update method, has been proposed to circumvent these issues by decoupling multi-spin interactions into two-spin interactions, thus accelerating the sampling process \cite{liu2017self}. Despite this, the general applicability of this method remains to be verified.~\\

\begin{figure} [t!]
\includegraphics[width=1\linewidth]{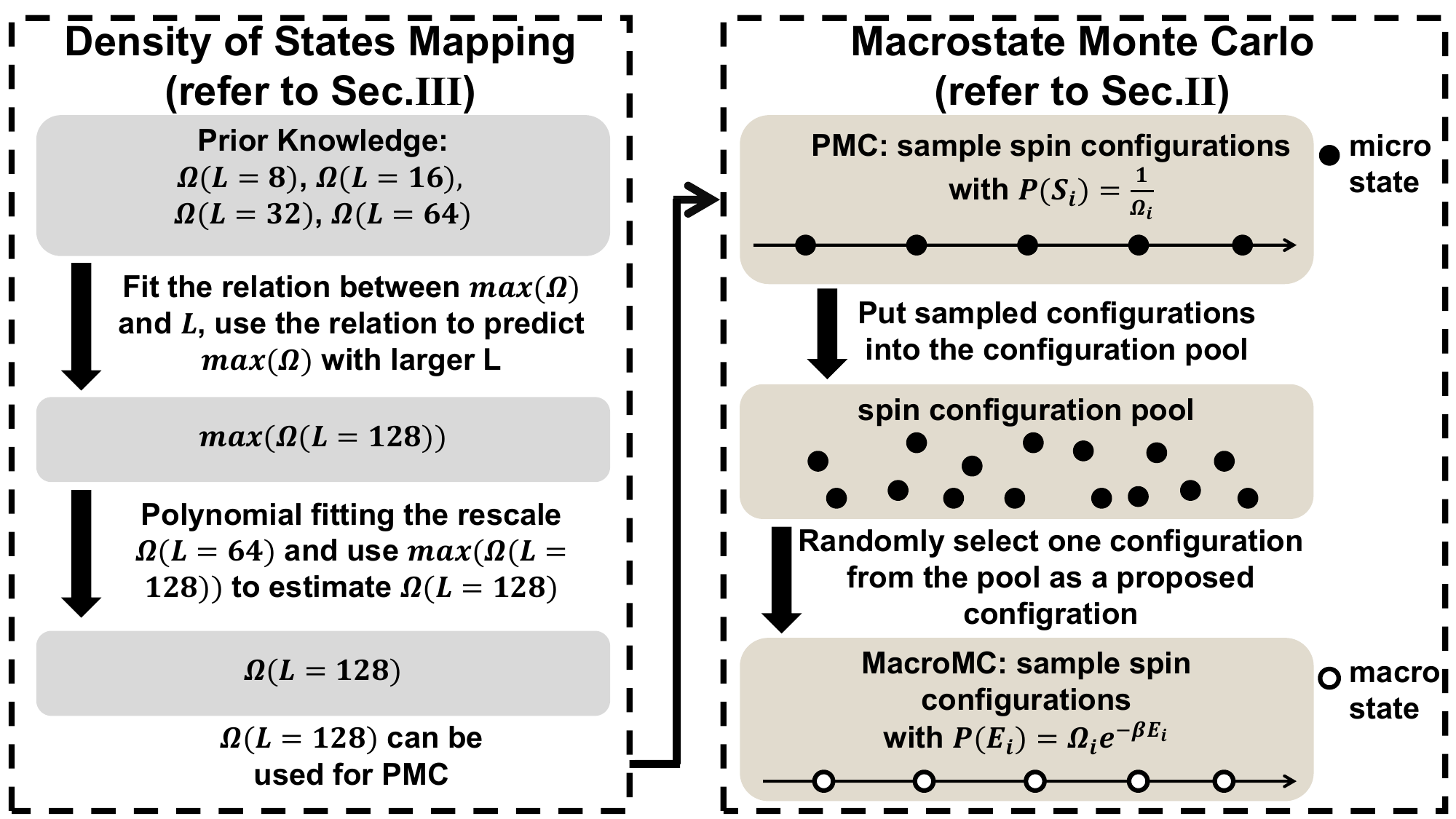}
\caption{The left and the right panel illustrates the density of states mapping method (Sec. \ref{sec:density_state_curve_mapping}) and the Macrostate Monte Carlo method (Sec. \ref{sec:macrocanonical_monte_carlo}), respectively. The mapping method comprises two phases. First, the relationship between the system size \(L\) and the maximum density of states for each respective \(L\) is established through fitting. The fitted function is then utilized to estimate the maximum density of states for larger systems. Second, the density of states for the largest system size already known is rescaled to a range between 0 and 1. A polynomial fit is then applied to model the rescaled density of states curve. Utilizing the previously estimated maximum density of states, this curve is then mapped back to its original scale, enabling accurate extrapolation of the density of states for larger systems. The Macrocannonical Monte Carlo (MacroMC) consists of two parts. The first part, Proposal Monte Carlo (PMC), assigns an occurrence probability to each state that is the inverse of the estimated density of states, with proposed spin configurations generated through random spin flips. The spin configurations sampled by PMC are stored in a spin configuration pool for use in the second part of the algorithm. The second part, MacroMC, assigns an occurrence probability to each state based on the Boltzmann probability of the corresponding macrostate, with proposed spin configurations randomly selected from the spin configuration pool.
}
\label{fig:fig1}
\end{figure}

To address the challenges of low sampling efficiency and high sample autocorrelation, this study proposes a general Monte Carlo method designed to enhance the efficiency of energy space sampling. In contrast to the traditional Metropolis-Markov Chain Monte Carlo (MMCMC) method, where each state represents a specific spin configuration, our method considers each state as a collection of all spin configurations with a specific energy. We called such a state as a ``macrostate", and hence named the approach as the Macrostate Monte Carlo or MacroMC. The key structure of the algorithm is shown on the right panel of Fig. \ref{fig:fig1}. In this method, the acceptance rate is changed from the conventional \(e^{- \beta(E_f-E_i)}\) to 
$e^{- \beta(E_f-E_i)}\Omega(E_f)/\Omega(E_i)$, where $\beta$ is the inverse temperature, $\Omega(E)$ is the density of states - the number of possible microstates (spin configurations) for a macrostate with energy $E$, while $E_i$ and $E_f$ is the initial and final energy, respectively. The introduction of the densities of states significantly increases the random walker's step size in the energy space, facilitating efficient sampling from the entire configuration space. Importantly, the required density of states \(\Omega\) does not necessitate exact values; sufficiently accurate estimates can be obtained via the Wang-Landau algorithm \cite{wang2001efficient}, the transition matrix Monte Carlo \cite{wang2002transition} and the ratio methods \cite{bi2015monte}. In the MacroMC algorithm, proposed spin configurations are not generated through local or cluster updates of the previous spin configuration but are randomly selected from a pool of spin configurations. This ensures the independence of successive spin configurations and allows for the rapid generation of a large number of trail configurations through parallel processing. The details of the algorithm, exemplified by the Ising model, is elaborated in Sec. \ref{sec:macrocanonical_monte_carlo}, along with its application to the 10-state Potts model, EA model, and the FFXY model. ~\\

The efficiency of MacroMC method relies on the estimation of density of states, among which the Wang-Landau algorithm is renowned for its high precision and ease of implementation \cite{wang2001efficient}. A plethora of studies aimed at optimizing the Wang-Landau algorithm to enhance its efficiency has been carried out \cite{belardinelli2007fast,caparica2012wang,vogel2013generic,koh2013dynamically,caparica2014wang,barash2017control,hayashi2019efficient,dai2020wang}. However, these optimizations have not overcome the limitation of the Wang-Landau algorithm, where the sampling time grows exponentially with the number of states. This problem limits in-depth research on critical phenomena in large systems. To address this issue, we introduce a mapping method that extrapolates the density of states from a small system to a large system without additional sampling. The method is illustrated in the left panel of Fig. \ref{fig:fig1} and a detailed discussion of the method and its effects is presented in Sec. \ref{sec:density_state_curve_mapping}. ~\\

\section{Macrostate Monte Carlo}
\label{sec:macrocanonical_monte_carlo}

\subsection{The method}
\label{subsec:method}
The algorithm consists of two types of Monte Carlo steps, as illustrated in the right panel of Fig. \ref{fig:fig1}. The first type is named Proposal Monte Carlo (PMC) in our algorithm. It is a normal flat energy histogram Monte Carlo sampling, where each state in the Markov chain represents a specific spin configuration. PMC is utilized to generate the proposed spin configurations, ensuring each macrostate has an equal probability of occurrence of \(1/M\), where \(M\) is the total number of macrostates. The occurrence probability of any microstate within a macrostate is thus \(1/(M\Omega(E(s)))\), where $s$ is the configuration of a microstate. During the updates, a site is randomly selected, and a spin flip is accepted with a probability of $R_{p} = \Omega(E(s_i))/\Omega(E(s_f))$. More discussion about the detailed balance of PMC can be found in the Appendix \ref{sec:Detailed_balance_of_proposal_monte carlo}. The spin configurations generated by PMC are then stored in a spin configuration pool, awaiting use by the next step.~\\

The second Monte Carlo step is MacroMC, which is used to reject or accept spin configurations sampled by PMC. In MacroMC, each macrostate represents an ensemble of all spin configurations at a specific energy level, thus the occurrence probability of each such state is the product of the density of states and the Boltzmann weight $W(E(s))=e^{-\beta E(s)}$, i.e. \(\Omega(E(s))W(E(s))\). During MacroMC updates, a spin configuration is randomly selected from the spin configuration pool and accepted with a probability of \(R_{m}\). To achieve detailed balance, the acceptance rate \(R_{m}\) must satisfy the following condition:
\begin{equation}
R_{m} = \frac{\Omega(E(s_{2}))W(E(s_{2})) T_{21}}{\Omega(E(s_{1}))W(E(s_{1})) T_{12}},
\label{eq1}
\end{equation}
where \(T_{12}\) represents the probability that the Markov chain is in the macrostate \(E_1\) at time \(t\) and transitions to the macrostate \(E_2\) at time \(t+1\),
\begin{equation}
T_{12} = P(E_{2}|E_{1}) = \frac{P(E_{1} \cap E_{2})}{P(E_{1})}.
\label{eq2}
\end{equation}

Unlike MMCMC, the proposed spin configurations in MacroMC are not obtained through local or cluster flips from the previous spin configuration but are randomly selected from the spin configuration pool, ensuring no correlation between them. Therefore, \(P(E_{1} \cap E_{2}) = P(E_{1}) \times P(E_{2})\). Since each macrostate in the spin configuration pool has an equal number of spin configurations, \(T_{12} = P(E_{2}) = 1/M\), and similarly, \(T_{21} = 1/M\). The expression for the acceptance rate \(R_{m}\) is then reduced to
\begin{equation}
R_{m} = \frac{\Omega(E(s_{2}))W(E(s_{2}))}{\Omega(E(s_{1}))W(E(s_{1}))}.
\label{eq3}
\end{equation}~\\

Whether in single update MCMC or cluster update MCMC, the proposed spin configuration at the simulation time \(t+1\) is generated by modifying the spin configuration at time \(t\), inherently introducing a high correlation between spin configurations. This correlation causes strong trends in the time series of measured quantities, preventing the random walker from efficiently exploring the entire configuration space in a short period. For the 2D Ising model with \(L=128\), single update MCMC stays within a narrow energy range for a long simulation time, whereas cluster update mitigates this trend, allowing for random walks across a broader energy range, but still lacking efficiency. In our algorithm, the proposed spin configuration appears with equal probability across all energy intervals and is uncorrelated with the spin configuration at the previous time time, offering the opportunity to explore a larger space more rapidly with much less autocorrelation. In this aspect, our algorithm surpasses previous algorithms in the step of proposing spin configurations.~\\

When sampling at a specific temperature using the usual MCMC, the outcome of Boltzmann distribution comes from a competition between the disorder induced by random spin flips in proposed spin configurations and the order induced by Boltzmann acceptance rates in accepted spin configurations. In our algorithm, although proposed spin configurations do not lead to disorder, the acceptance rate \(R_{m}\) is governed by both the density of states \(\Omega\) and the Boltzmann weight \(W\). As energy decreases, the density of states exponentially decreases, and the Boltzmann weight exponentially increases. When the energy of the proposed spin configuration at time \(t+1\) is lower than that at time \(t\), the increase in Boltzmann weight significantly surpasses the decrease in density of states. Conversely, when the energy of the proposed spin configuration increases, the increase in density of states greatly exceeds the decrease in Boltzmann weight. The interplay between density of states and Boltzmann weight leads to sampling outcomes comparable to those of MCMC. Furthermore, by inducing density of states, our algorithm can achieve a higher acceptance rate even when there is a significant energy difference between the proposed spin configuration at time \(t+1\) and the accepted spin configuration at time \(t\), allowing our algorithm to move greater distances in the energy space at one sweep, and resulting smaller autocorrelation times.~\\

In fact, when multicanonical Monte Carlo uses \(W = \beta E(S_{i}) + \ln(1/\Omega_{i})\) to rewrite the Boltzmann weight, the weight of each microstate becomes \(1/\Omega_{i}\) similar to PMC. Subsequently, multicanonical Monte Carlo employs a reweighting technique to estimate physical quantities under the Boltzmann weight, whereas we use MacroMC for this estimation. However, reweighting in multicanonical Monte Carlo involves calculations with large numbers in the exponent, which can easily lead to computational overflow. Although this can be circumvented by scaling both the numerator and the denominator, it introduces additional complexity. The problem is automatically avoided in our algorithm since it does not involve any reweighting.~\\

\subsection{Performance in 2D Ising model}
\label{subsec:performance_in_2D_ising_model}
We compared the autocorrelation of sampling at the phase transition temperature (\(T_c\approx 2.267\)) for the 2D Ising model on a square lattice using single update MCMC, cluster update MCMC, and MacroMC. The Hamiltonian of the Ising model reads
\begin{eqnarray}
H = -\sum_{\langle i, j\rangle} S_i S_j,
\label{eq4}
\end{eqnarray}
where the sum $\left< i,j\right>$ is over the nearest neighbors, and $S_i\in\{-1,1\}$. Unless otherwise specified,  we consider a system of $N=L\times L=128\times 128$ spins. The autocorrelation at the simulation time $t$ can be quantified by
\begin{equation}
A(E,t) = \frac{\langle E(t_{0})E(t_{0} + t) \rangle - {\langle E(t_{0}) \rangle}^{2}}{\langle {E(t_{0})}^{2} 
\rangle - {\langle E(t_{0}) \rangle}^{2}},
\label{eq5}
\end{equation}
where $t_0$ is the initial time. ~\\

\begin{figure} [t!]
\centering
\hspace{0.25cm} 
\includegraphics[width=0.45\linewidth]{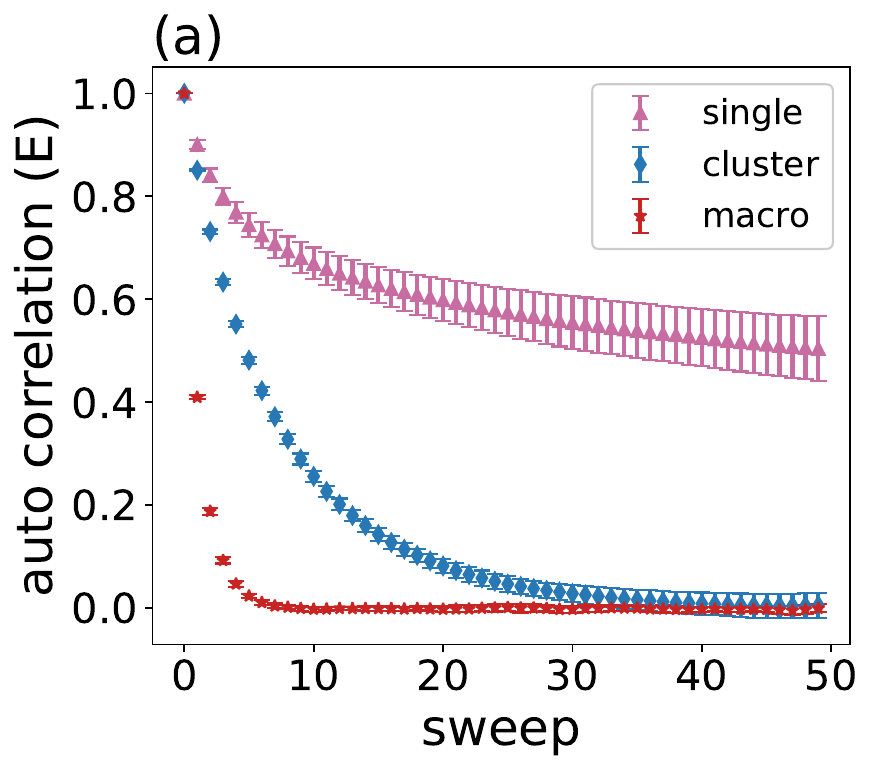}
\hspace{0.2cm} 
\includegraphics[width=0.45\linewidth]{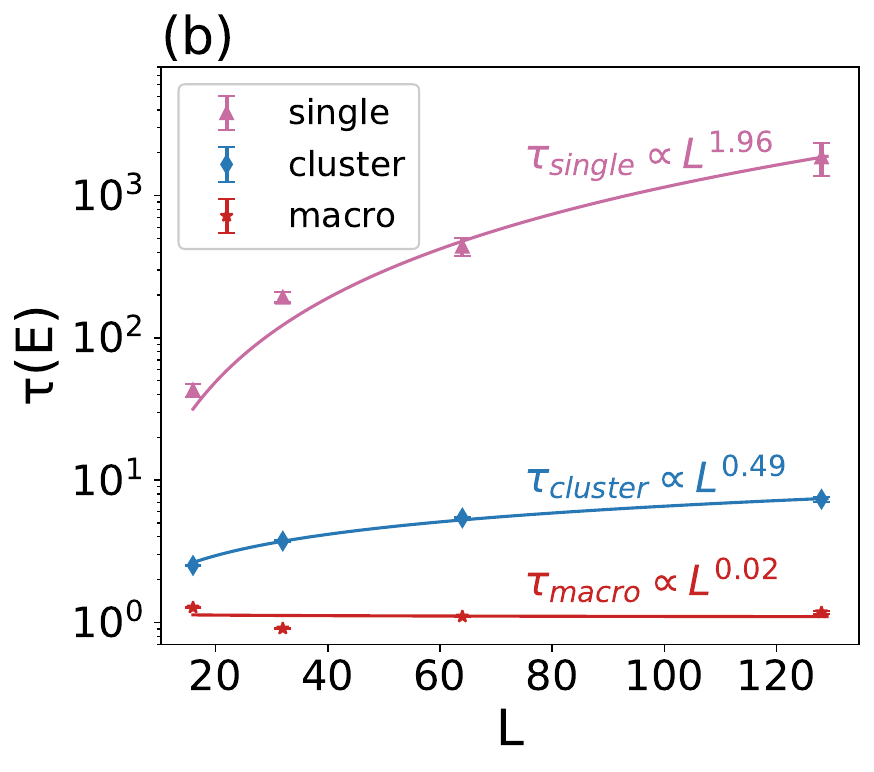}
\centering
\includegraphics[width=0.49\linewidth]{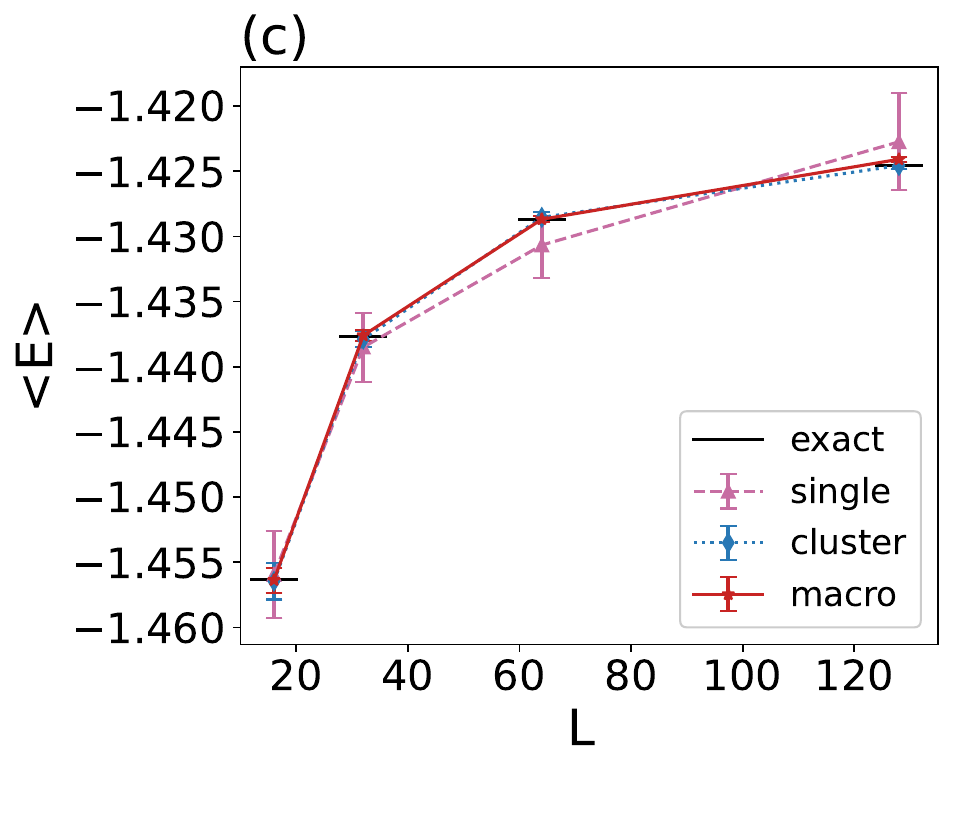}
\includegraphics[width=0.485\linewidth]{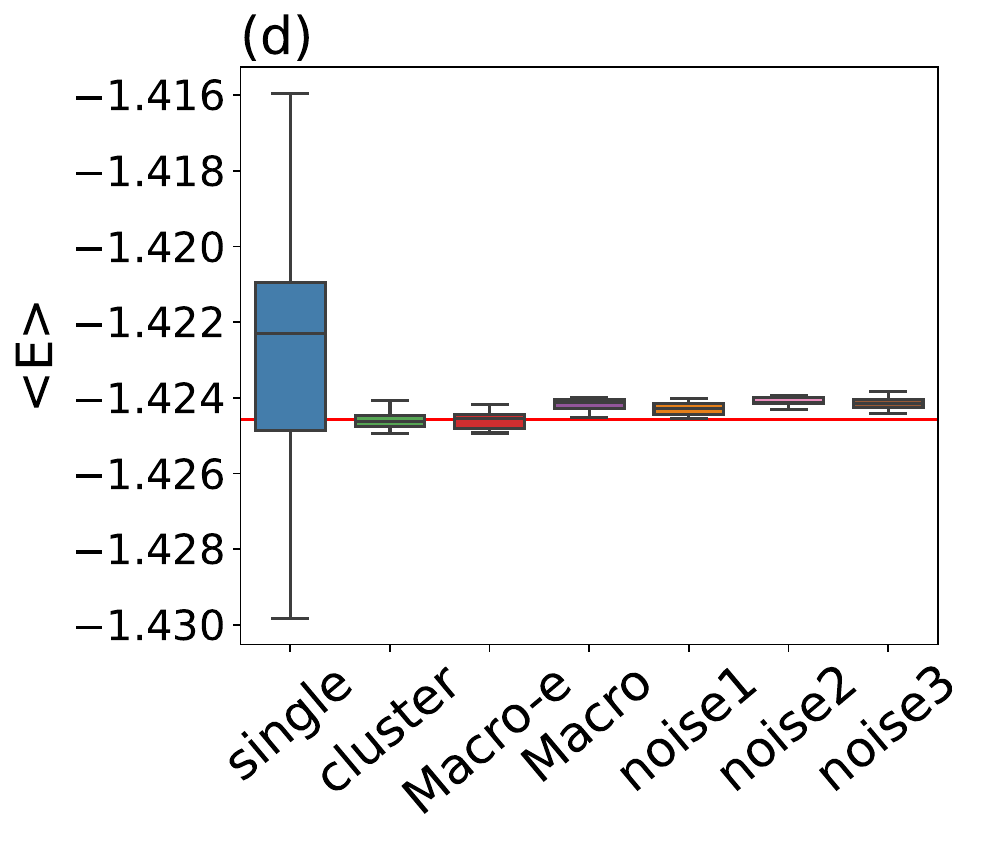}
\caption{(a) The figure shows the autocorrelation of the 2D Ising model at \(L=128\) during phase transition, when sampled using single update MCMC, cluster update MCMC, and MacroMC. The autocorrelation for MacroMC is significantly lower than that for single update MCMC and cluster update MCMC. (b) The autocorrelation time of the three algorithms for different system sizes, where the autocorrelation time of MacroMC almost does not increase with \(L\). (c) The figure shows the measured energy at the phase transition temperature for the three algorithms. Both MacroMC result and cluster update MCMC result are close to the exact values and have small errors. (d) The figure presents the results of Boltzmann sampling for the \(L=128\) system using MacroMC with exact density of states (Macro-e), estimated density of states without noise (Macro) and estimated density of states with noise (noise1 to noise3). The box in the plot represents the  interquartile range (25th to 75th percentiles) of the data with the horizontal line inside denotes the median. The end of the upper and lower whisker marks the maximum and minimum values of the data respectively. Despite the large noise, the results of MacroMC are close to the exact values (represented by the red line), demonstrating the robustness of MacroMC. }
\label{fig:fig2}
\end{figure}

We compare our algorithm with the standard single-flip Monte Carlo and cluster-flip Monte Carlo. As shown in Fig. \ref{fig:fig2}(a), our algorithm reduces the autocorrelation to below 0.4 after just one sweep, significantly faster than a single update and also outperforming cluster update. Note that, our algorithm does not reduce autocorrelation to almost zero in one sweep. This is because each marcostate formed from spin configurations with the same energy in the pool are equally probable, and those with a low Boltzmann probability will be rejected easily and multiple rejections can result in a constant measured value in subsequent time steps, thus introducing autocorrelation.
Fortunately, the acceptance rate of our algorithm does not decrease with an increased system size. ~\\

We fitted the autocorrelation function for each algorithm at different system sizes using an exponential function \(A = k_1 \cdot \exp(-t/\tau)\), where \(t\) represents the sweep, \(k_1\) and \(\tau\) are fitting parameters, and \(\tau\) is the autocorrelation time to be extracted. For the cluster update Monte Carlo and MacroMC, the autocorrelation function drops to around 0 within 100 sweeps, so we fitted the short-time autocorrelation function in range \(t \in [0, 100]\). However, for the single update Monte Carlo, more sweeps are required for convergence, so we fitted the long-time autocorrelation function in range \(t \in [0, 1000]\). The fitting results are shown in Figure \ref{fig:fig2}(b). One can see that the autocorrelation time of the MacroMC method remains low and has a small size dependence, i.e. $\tau\sim L^{0.02}$. On the other hand, the autocorrelation time for single and cluster update methods show a significant exponential increase with the system size, where $\tau\sim L^{1.96}$ and $\tau\sim L^{0.49}$, respectively. The scaling exponent obtained for the single update method is consistent with the previous study \cite{nightingale1996dynamic}.~\\

In fact, we could also compare our algorithm with Replica Exchange Monte Carlo. However, the performance of Replica Exchange Monte Carlo is sensitive to algorithm parameters such as exchange frequency and the temperature difference between Monte Carlo processes. This sensitivity can introduce bias in the comparison results. Additionally, as shown in Fig. \ref{fig:fig2}, our algorithm is significantly better than the cluster-flip Monte Carlo, which proves that our algorithm is valuable.~\\

In our algorithm, PMC does not need to wait for MacroMC to reject or accept an old proposed spin configuration before proposing a new one. Therefore, PMC and MacroMC can be processed independently, allowing us to process multiple PMCs in parallel, thus getting a large number of proposed spin configurations within the same wall time. When estimating physical quantities using MacroMC, the only time-consuming operation is calculating the acceptance rate. This makes MacroMC nearly cost-free in time and computational resources.~\\

We also compared the accuracy of the observable measured by single update, cluster update, and MacroMC methods. We estimated the energy of the 2D Ising model with \(L = \{16, 32, 64, 128\}\) at \(T = 2.267\). For single and cluster updates, each \(L\) underwent 10 independent Monte Carlo runs, and each run performed \(10^{6}\) measurements. 
In our algorithm, to maximize the advantages of PMC and MacroMC, we conducted 10 independent PMCs in parallel for each \(L\), obtaining a total of \(10^{7}\) proposed spin configurations. Then, \(10^{6}\) proposed spin configurations were randomly drawn from the pool for  MacroMC and the process was repeated 10 times. To prevent spin configurations generated by PMC from being completely rejected in MacroMC sampling, we set the density of states that approaches zero at \(T=2.267\) to infinity. Figure \ref{fig:fig2}(c) shows our results, where error bars represent the standard deviation of 10 energy measurements. The exact energy values were obtained using the exact density of states \cite{wang2001efficient}. The autocorrelation in single updates led to larger errors and discrepancies from the exact values. Both cluster update and MacroMC nearly eliminated autocorrelation in the 2D Ising model, resulting in much smaller errors compared to single updates. The results proves the effectiveness of our algorithm.~\\

Given our algorithm's reliance on the density of states, a legitimate concern is whether PMC can sample each macrostate with equal probability and estimate accurate physical quantities through MacroMC when the estimated density of states significantly deviates from the exact one. We added three different types of noise to the estimated density of states of the Ising model with \(L=128\): noise1 being a normal distribution $\mathcal{N}(0,1)$, \({\text{noise2}} = \sin(E)\), and \({\text{noise3}} = \sin(0.0008E)\). We then used PMC and MacroMC to sample from these noisy densities of states and measured the energy. Figure \ref{fig:fig2}(d) shows the sampling results for different Monte Carlo algorithms when \(L = 128\). The MacroMC result using the estimated density of states without noise is slightly larger than the exact solution (labelled as MacroMC in the figure). This may attribute to the discrepancy in the PMC sampling from the estimated density of states instead of the inadequacy of the MacroMC algorithm itself (also see the discussion in Appendix \ref{appB}). On the other hand, if the exact density of states is used, PMC sampling is unnecessary in this case. The exact sampling result of the configuration pool, where each macrostate has the same occurrence probability, is automatically known. The average energy obtained in this way (labelled as Macro-e in Fig. \ref{fig:fig2}(d)) agrees well with the exact solution. Alternatively, one can also improve the accuracy using more sophisticated methods such as combining PMC sampling with replica-exchange method \cite{vogel2014scalable}. Nevertheless, the three noise results fluctuate slightly around the MacroMC result using the estimated density of states, indicating that the MacroMC simulation is not affected by noise significantly. Even with the most locally fluctuating \({\text{noise1}}\), MacroMC achieves results far better than those of single update and comparable to cluster update, demonstrating our algorithm's robustness. More details about the noise and the MacroMC result can be found in Appendix \ref{sec:Robustness_of_MacroMC}.


When there is a difference between the estimated and exact density of states, PMC will no longer sample each macrostate with equal probability. In other words, the resulting energy histogram will not be flat but will show trends related to the errors in the estimated density of states. Fortunately, this does not bring significant issues in our algorithm. Specifically, if the estimated density of states \(\Omega_{\text{estimate}}(E_i)\) for an energy \(E_i\) is lower than the exact density of states \(\Omega_{\text{exact}}(E_i)\), the acceptance rate \(R_p\) of PMC will increase the sampling probability for this macrostate. However, when MacroMC samples proposed spin configurations belonging to \(E_i\), the acceptance rate \(R_m\) will decrease the acceptance probability for this macrostate, ultimately yielding accurate results. We provide a detailed proof in the Appendix \ref{sec:Sampling_with_non-exact_density_of_state} that MacroMC can give a Boltzmann distribution using an estimated density of states. In practice, it is unlikely to have the estimated density of states as adversely affected as those by noise1, noise2, and noise3. Through the ratio method, we can easily obtain an estimated density of states with local fluctuations less than 0.1 compared to the exact density of states. Using the Wang and Landau algorithm, we can obtain an even more accurate estimation of the density of states.~\\

\subsection{Generalization to different models}
\label{subsec:application_on_other_models}

We applied our algorithm to three representative models, namely the FFXY, 10-state Potts, and the EA models. The Hamiltonians of these models are defined as follows,
\begin{eqnarray}
H_{\rm{FFXY}} &=& -J \sum_{\langle i, j \rangle} \cos(\theta_i - \theta_j - A_{ij}),\\
H_{\rm{Potts}} &=& -J \sum_{\langle i, j \rangle} \delta (\sigma_i, \sigma_j ),\\
H_{\rm{EA}} &=& -\sum_{\langle i, j \rangle} J_{ij} S_i S_j,
\end{eqnarray}
where $J$ and $J_{ij}$ are the coupling constants, with $J$ set to 1 and $J_{ij}$ a random variable that can be $1$ or $-1$. The sum $\langle i,j \rangle$ is over nearest neighbors. In the FFXY model, $\theta_i$ represents the angle of the spin at site $i$, and $A_{ij}$ denotes the magnetic flux distribution such that the sum of $A_{ij}$ over each plaquette is $\pi$. In the Potts model, $\sigma_i$ denotes the spin at site $i$, which can take positive integer values in $\{1, 2, 3, \ldots, 10\}$, and $\delta (\sigma_i, \sigma_j)$ is the Kronecker delta. In the EA model, $S_i$ denotes the spin at site $i$, taking values $\pm 1$.~\\

The FFXY model is an extension of the classical XY model, and its rich phase diagram has attracted attention due to its correspondence with experimental observations in magnetic thin films and Josephson junction arrays \cite{halsey1985topological,kawamura1999successive}. This model features checkerboard chirality, which leads to a competition between Ising-like phase transitions and Berezinskii-Kosterlitz-Thouless (BKT) transitions \cite{teitel1983phase,olsson1995two,hasenbusch2005multicritical}. At high temperatures, the system is in a disordered phase. As the temperature decreases, the system first undergoes an Ising-like phase transition and then, at a lower temperature, experiences a BKT transition. The Ising-like transition is related to the spontaneous breaking of the plaquette checkerboard chiral symmetry, while the BKT transition is related to the dissociation of vortex-antivortex pairs. An unusual feature is that the temperature gap between the Ising-like transition and the BKT transition is only about 0.02 in units of \( J/k_B \) for finite system sizes. During a low-temperature Monte Carlo simulation, the four spins in a plaquette can freeze easily due to the magnetic flux. To shift to another spin configuration with a different spin pattern, a spin configuration needs a large number of Monte Carlo steps to accumulate small changes in each step, leading to large autocorrelations in spin configurations.~\\

The finite-state Potts model is an extension of the Ising model and is known for its complex phase transition behaviors. It provides a statistical description of various real systems, such as alloy phase transitions and biological population behaviors \cite{sahni1983kinetics,rozikov2022gibbs,tikare1998application,miodownik2002review,laanait1986phases,mortazavi2013study,kotecky1990q}. At low temperatures, the model undergoes metastable tunneling, where the system can switch between local energy minima. This feature can also describe multiple metastable crystal structures in alloy systems at low temperatures and multiple metastable biological populations in ecological systems under low environmental perturbations. As the temperature increases to the transition temperature, the system exhibits transitions between energy peaks and undergoes a first-order phase transition \cite{wu1982potts,baxter1973potts,fukugita1989correlation,nardi2019tunneling}, which can also describe dramatic switching between multiple states in real systems under severe environmental disturbances. During a Monte Carlo simulation at the transition temperature, the Markov chain walker at one peak needs to overcome a large energy barrier to jump to another peak, resulting in large autocorrelation.~\\

The EA model is a typical spin glass model and is also an extension of the Ising model. It is defined by random interaction strengths between spins and can be used to explain the glassy behavior of some magnetic materials and protein folding \cite{katzgraber2007finite,thirumalai1996dynamics}. Because this randomness, the EA model exhibits a glassy phase at low temperatures, which is distinct from the paramagnetic phase \cite{zhao2011methods,moreno2003finding}. The glassy phase is characterized by a lack of long-range order and a highly disordered spin structure, leading to slow dynamics and a complex energy landscape. During a Monte Carlo simulation at low temperatures, a spin can be easily frozen by surrounding random interaction strengths \(J_{ij}\), which is similar to the FFXY model and causes large autocorrelation.~\\

The three models have complicated energy landscapes and strong autocorrelations in the MCMC simulations. While each has developed its own cluster update techniques, a simple, general, and efficient algorithm has yet to be established. Recall that the proposed spin configuration in MacroMC is randomly selected from a configuration pool. The spin configurations in the pool are sampled by PMC in which the complication of energy landscapes is ignored. Therefore, the proposed spin configurations in MacroMC have much less autocorrelation. To demonstrate the performance of our algorithm, we sampled the FFXY model, 10-state Potts model, and the EA model for \(L = 64\) and estimated their autocorrelation times at \(T=0.45\), \(T=0.7\), and \(T=1\), respectively.~\\

\begin{figure} [t!]
\includegraphics[width=0.47\linewidth]{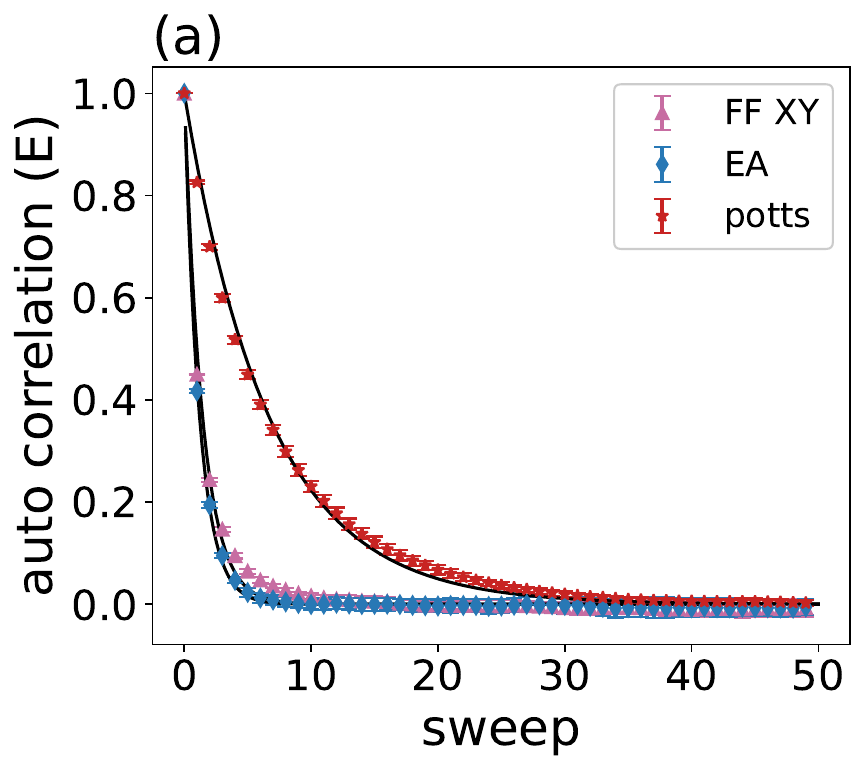}
\includegraphics[width=0.5\linewidth]{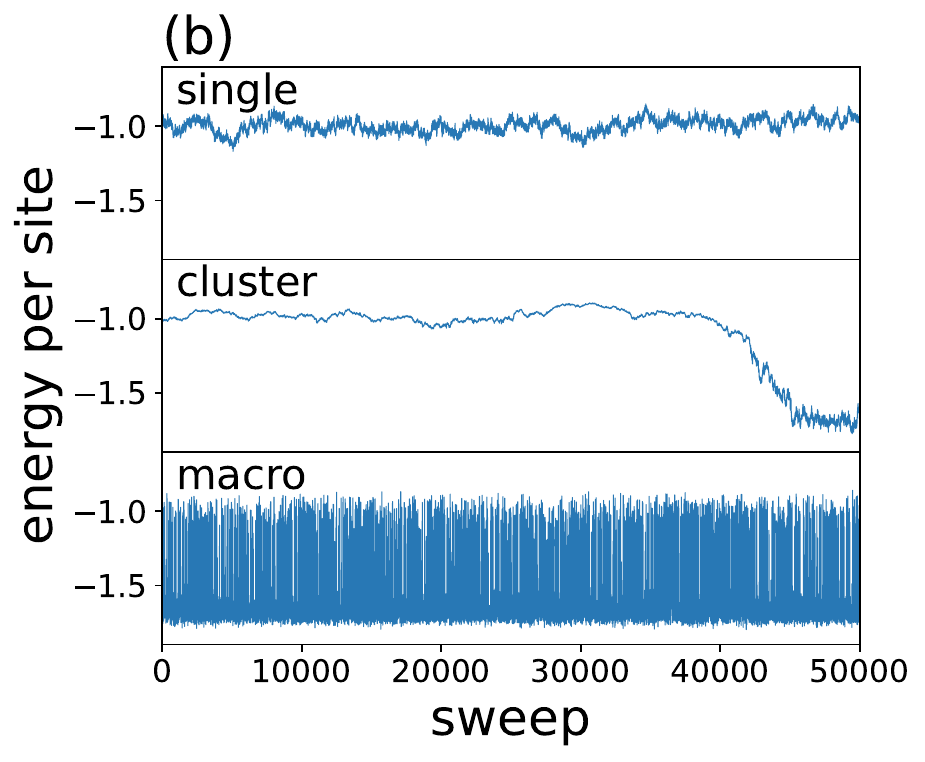}
\includegraphics[width=0.47\linewidth]{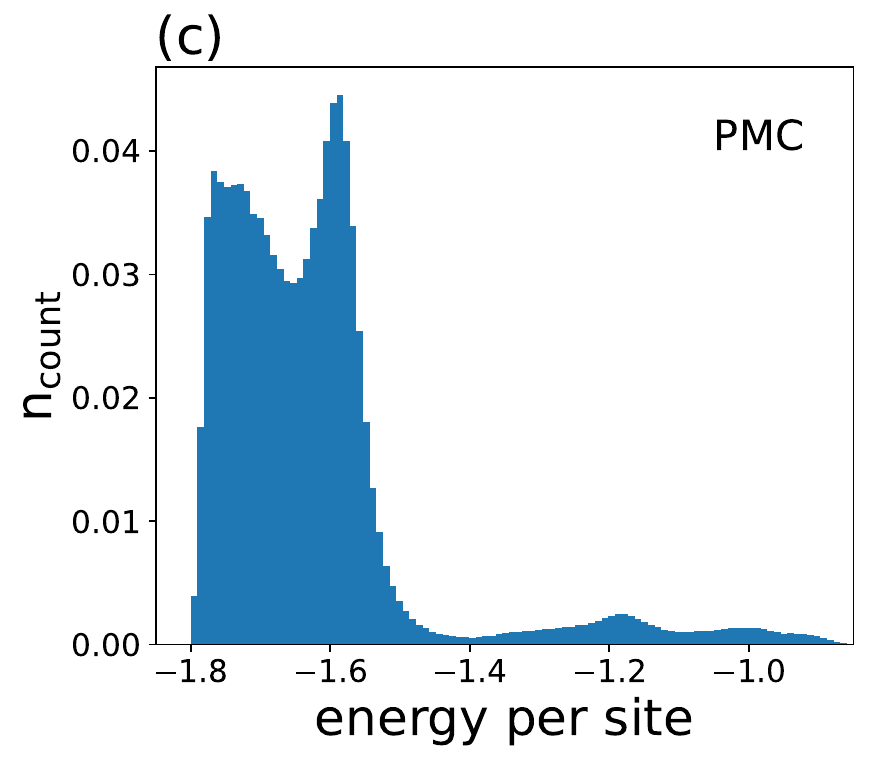}
\includegraphics[width=0.47\linewidth]{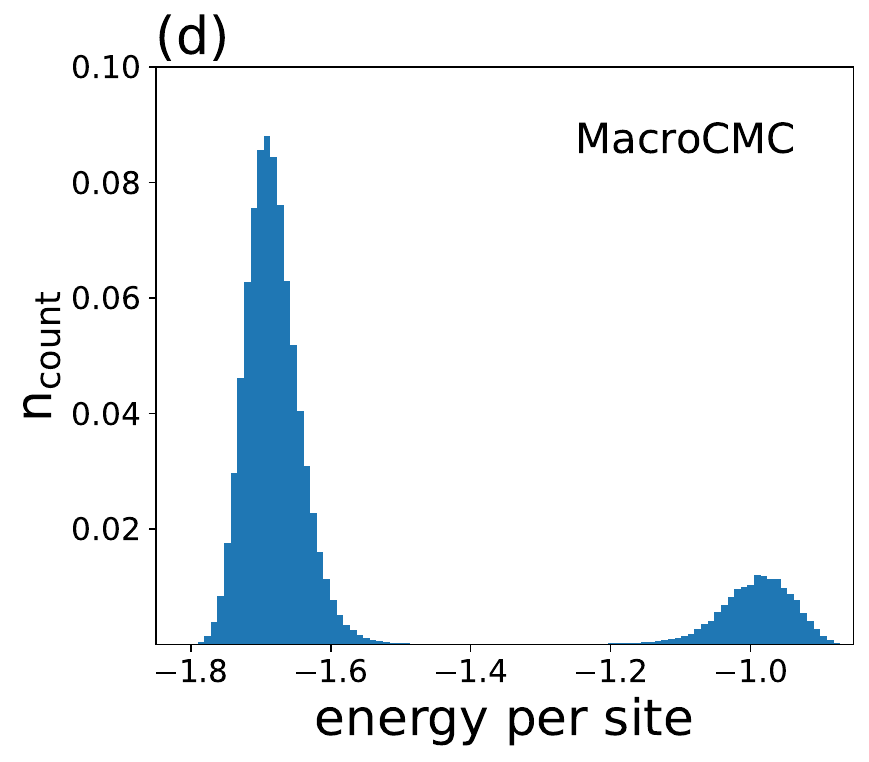}
\caption{(a) The figure shows the autocorrelation of MacroMC near the phase transition temperature in the FFXY, the EA, and the 10-state Potts models, demonstrating low autocorrelation times across all three models. (b) The figure traces the movement of a random walker in the energy space during the sampling process in the \(L=64\) 10-state Potts model using single update, cluster update plus replica exchange, and MacroMC. The single update algorithm is completely trapped in a local optimum, cluster update plus replica exchange escapes the local optimum after approximately \(4 \times 10^4\) sweeps, while MacroMC freely explores the entire energy space. (c) and (d) show the sampling results of PMC and MacroMC for the \(L=64\) 10-state Potts model, respectively. Even the PMC sampling distribution is non-flat, the MacroMC is still able to sample results consistent with the Boltzmann distribution.}
\label{fig:fig3}
\end{figure}

Figure \ref{fig:fig3}(a) shows that our algorithm achieves very low autocorrelation times across the three models, with \(\tau = 1.2, 1.4, 6.7\) for the EA, FFXY, and the Potts model, respectively. In Fig. \ref{fig:fig3}(b), we compare the sampling efficiency of single update, cluster update plus replica exchange, and our algorithm in the 10-state Potts model. The Potts model exhibits first-order phase transition and obtaining an accurate estimation of the density of states can be challenging. Over \(5 \times 10^4\) sweeps, the single update is entirely trapped within an energy region, while cluster update plus replica exchange succeeds in tunnelling only once, whereas our algorithm demonstrates frequent tunnelling. ~\\

As shown in Fig. \ref{fig:fig3}(c), significant differences between the estimated density of states and that obtained from the Wang-Landau algorithm in a non-flat energy histogram from PMC sampling. However, this does not prevent MacroMC from sampling a nice Boltzmann distribution in energy, as shown in Fig. \ref{fig:fig3}(d).~\\


\begin{figure} [t!]
\includegraphics[width=0.48\linewidth]{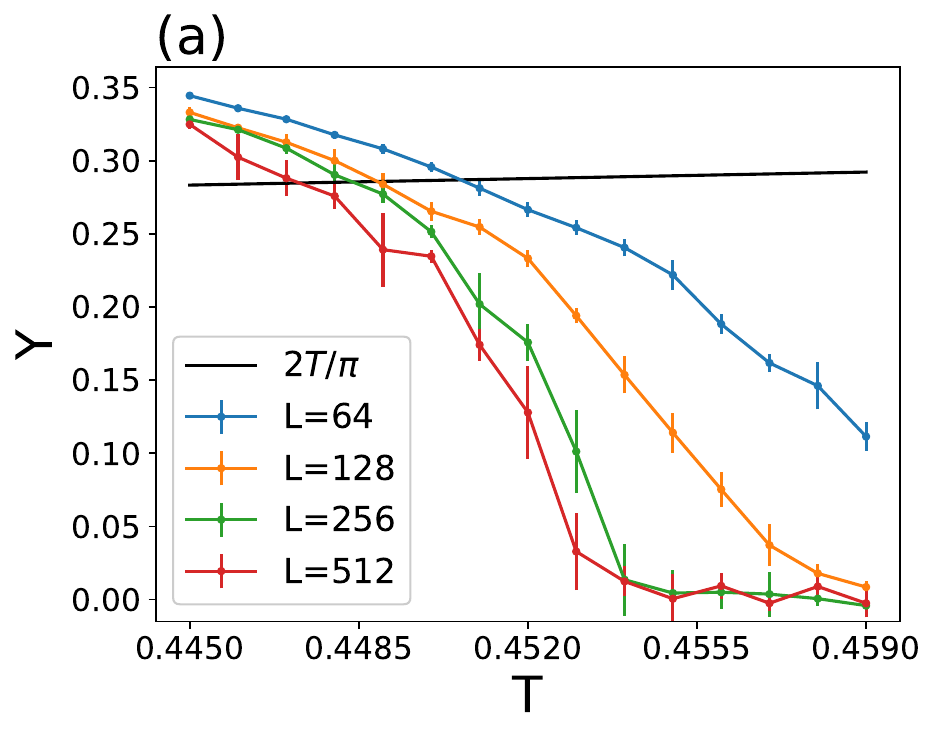}
\includegraphics[width=0.48\linewidth]{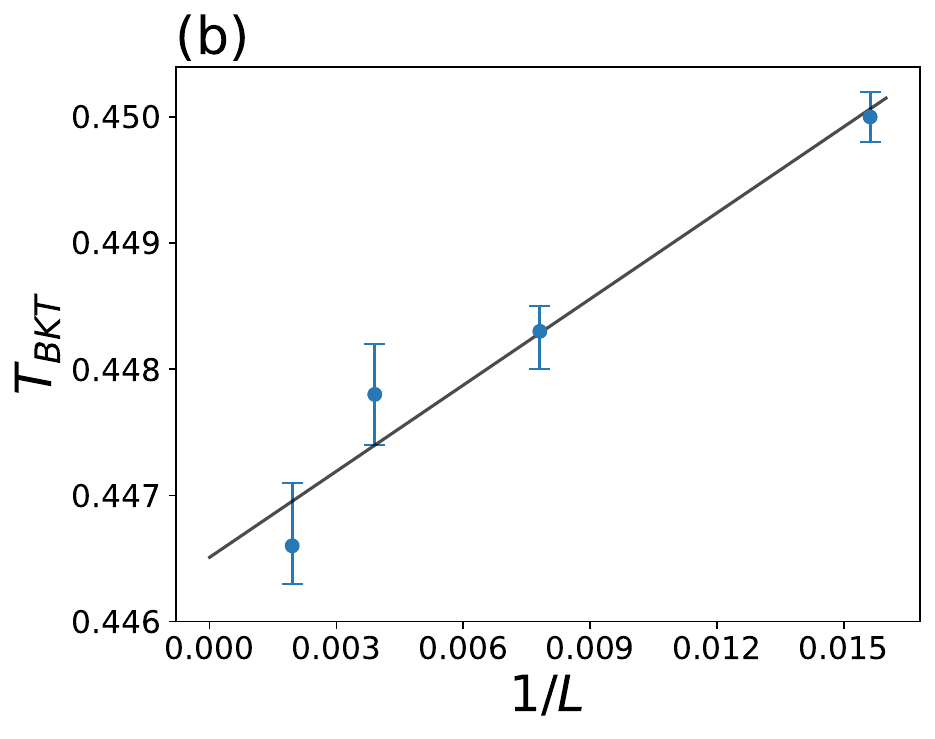}
\caption{(a) Helicity modulus (\(\Upsilon\)) of the FFXY model across different system sizes using MacroMC. Black line refers to \(\Upsilon = 2k_{B}T/\pi\), in which its intersection with the measured helicity modulus locates the BKT transition temperature. (b) The variation of the extracted \(T_{\text{BKT}}(L)\) with \(1/L\). The black line shows the linear fit to the data points, allowing for the estimation of \(T_{\text{BKT}}(L = \infty) \approx 0.4465\). }
\label{fig:fig4}
\end{figure}

Soichirou Okumura \textit{et. al.} employed the relaxation Monte Carlo method to sample the FFXY model and extensively studied the critical behaviour of the model\cite{okumura2011spin}. However, relaxation Monte Carlo still exhibits significant autocorrelations. We use MacroMCM to sample the FFXY model and determine the critical temperature of the BKT phase transition by calculating the helicity modulus $\Upsilon$, defined as follows:
\begin{equation}
\begin{aligned}
& \Upsilon_{\cos} = \left\langle \frac{1}{2}\sum_{i=1}^{L^{2}}{\cos{(\theta_{i} - \theta_{i-L})} + \cos{(\theta_{i} - \theta_{i+L})}} \right\rangle, \\
& \Upsilon_{\sin}^{2} = \left\langle \left( \frac{1}{2} \sum_{i=1}^{L^{2}} \left( \sin{(\theta_{i} - \theta_{i-L})} - \sin{(\theta_{i} - \theta_{i+L})} \right) \right)^{2} \right\rangle, \\
& \Upsilon_{y} = \frac{1}{L^{2}}\lbrack\Upsilon_{\cos} - \beta\Upsilon_{\sin}^{2}\rbrack,
\end{aligned}
\label{eq6}
\end{equation}
where \(i\) denotes the spin index, \(i-L\) and \(i+L\) is the index for the spin above and below the $i$th spin on the square lattice, respectively. The critical temperature \(T_{\text{BKT}}\) is located by the intersection of \(\Upsilon\) with the line \(\Upsilon = 2k_{\text{B}}T/\pi\), where \(k_{\text{B}}\) is the Boltzmann factor and is set to 1 for simplicity \cite{fisher1973helicity, nelson1977universal, canova2014kosterlitz}.~\\

We measure helicity modulus in the temperature range \(T\in [0.445,0.459]\) with an interval of 0.001. A total of \(10^{7}\) PMC steps are taken, and 5 MacroMCs are processed in parallel. Each MacroMC randomly selects \(2\times 10^{6}\) spin configurations to measure the physical quantities. Error bars are taken as the standard deviation of 5 measurements in the helicity modulus at a specific temperature. The sampling results from MacroMC, as shown in Fig. \ref{fig:fig4}(a), follow smooth curves across various system sizes even at such fine temperature intervals, suggesting the efficacy of MacroMC. Notably, \(T_{\text{BKT}}(L)\) decreases with increasing system size. 
We plotted \(T_{\text{BKT}}(L)\) as a function of \(1/L\) in Fig. \ref{fig:fig4}(b) and estimate \(T_{\text{BKT}}(L = \infty) \approx 0.4465 \pm 0.0003\), which is slightly higher than the values previously estimated using relaxation Monte Carlo $T_{\text{BKT}}\approx 0.4462$ \cite{okumura2011spin} and tensor network techniques $T_{\text{BKT}}\approx 0.4459$ \cite{song2022tensor}.~\\

\section{Mapping the density of states}
\label{sec:density_state_curve_mapping}
In this section, we introduce a scheme of estimating the density of states of a larger system from a smaller system. The scheme is inspired by two key observations. Firstly, the configuration space of smaller systems is substantially smaller than that of larger ones. The density of states in the small system can be obtain much more easily and accurately. Secondly, for a specific model, except some minor local details,  the change in the overall shape of the logarithmic of the density of states curve with respect to the energy is insensitive to the system size. This consistency allows us to infer the density of states of larger systems by analyzing the trends in the smaller ones.~\\

\begin{figure} [t!]
\includegraphics[width=0.47\linewidth]{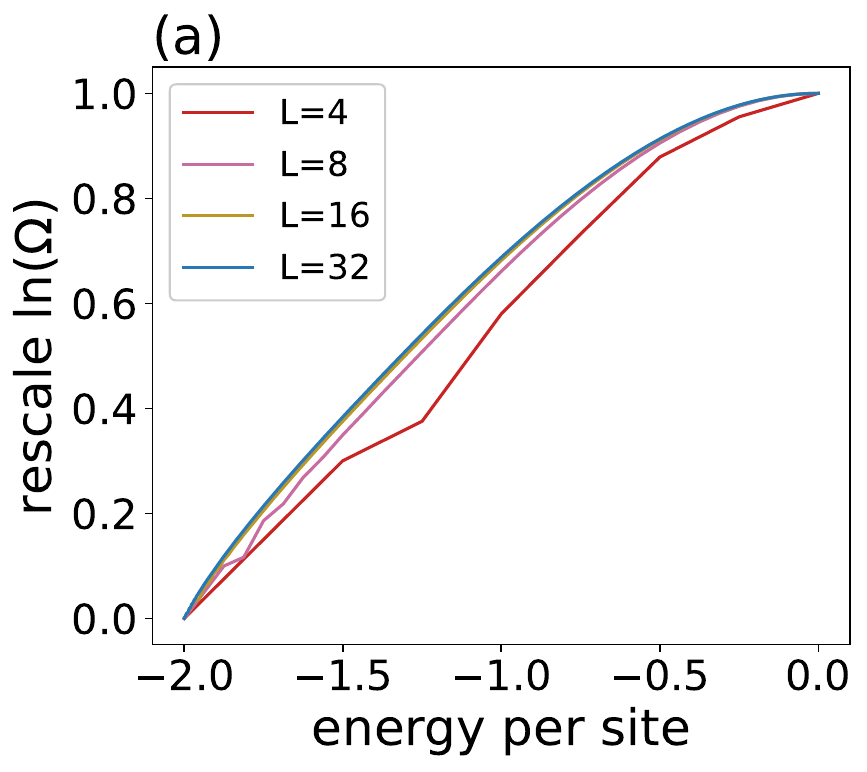}
\includegraphics[width=0.5\linewidth]{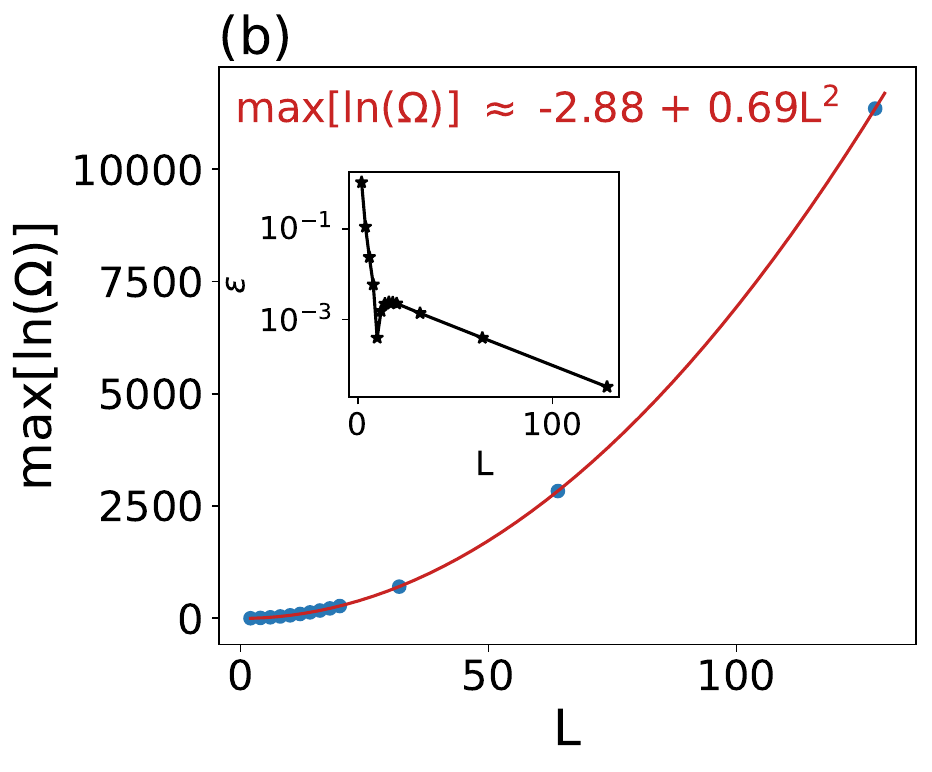}
\caption{(a) The rescaled density of states curves of the 2D Ising model for various system sizes. As \(L\) increases, the shapes of the curves tend to converge. (b) The relationship between the maximum logarithmic density of states and \(L\). The red curve shows the quadratic fit to the data. The inset shows the difference between the fitted and the exact values.}
\label{fig:fig5}
\end{figure}

For instance, in the 2D Ising model, the exact density of states is known, enabling the study of how the density of states curve evolves with increasing system size \cite{beale1996exact}. We first normalize the logarithmic density of states $\ln[\Omega(E)]$ to a scale within 0 to 1 by dividing it with $\ln[\Omega(E = 0)]$. As shown in Fig. \ref{fig:fig5}(a), the curves for small system sizes such as \(L=4\) and \(L=8\) are relatively rough and their shapes vary significantly. However, with increasing system size, the normalized logarithmic density of states rapidly converge to a unified curve, with the curves for \(L=16\) and \(L=32\) almost aligning with each other over the energy ranges of concern. This alignment suggests that the normalized logarithmic density of states of medium-sized systems closely approximates that of larger systems. The subsequent challenge involves mapping the normalized $\ln[\Omega(E)]$ back to the original density of states, necessitating the determination of the maximum density of states for each system size. As illustrated in Fig. \ref{fig:fig5}(b), the relationship between the maximum of the density of states and the system size can be accurately modelled by a quadratic function, with the error between the fitted and the exact values decreasing linearly as the system size increases for large enough system. This quadratic function facilitates the mapping of the normalized density of states for systems of any size back to their original densities of states.~\\

To illustrate our method of estimating the density of states without additional sampling, we present a specific example where the density of states for \(L=16\) is used to estimate the density of states for \(L=32\). We begin by estimating the density of states of the 2D Ising model for \(L=16\) using a sampling-based approach. The density of states is then rescaled to $[0, 1]$, and a polynomial fit is applied to the normalized density of states,
\begin{equation}
\ln\lbrack\Omega_{fit}(E)\rbrack = \sum_{i = 0}^{m}{k_{i}E_{i}},
\label{eq7}
\end{equation}
where \(m\) represents the number of terms, and \(k_i\) are the parameters to be determined from the fitting. Finally, with the quadratic function depicted in Fig. \ref{fig:fig5}(b), we map the density of states from \(L=16\) to \(L=32\) using
\begin{equation}
\ln\lbrack\Omega_{pre}(E)\rbrack = ( - 2.88 + 0.69 \times 32^{2}) \times \sum_{i = 0}^{m}{k_{i}E_{i}}.
\label{eq9}
\end{equation}

\begin{figure} [t!]
\includegraphics[width=0.46\linewidth]{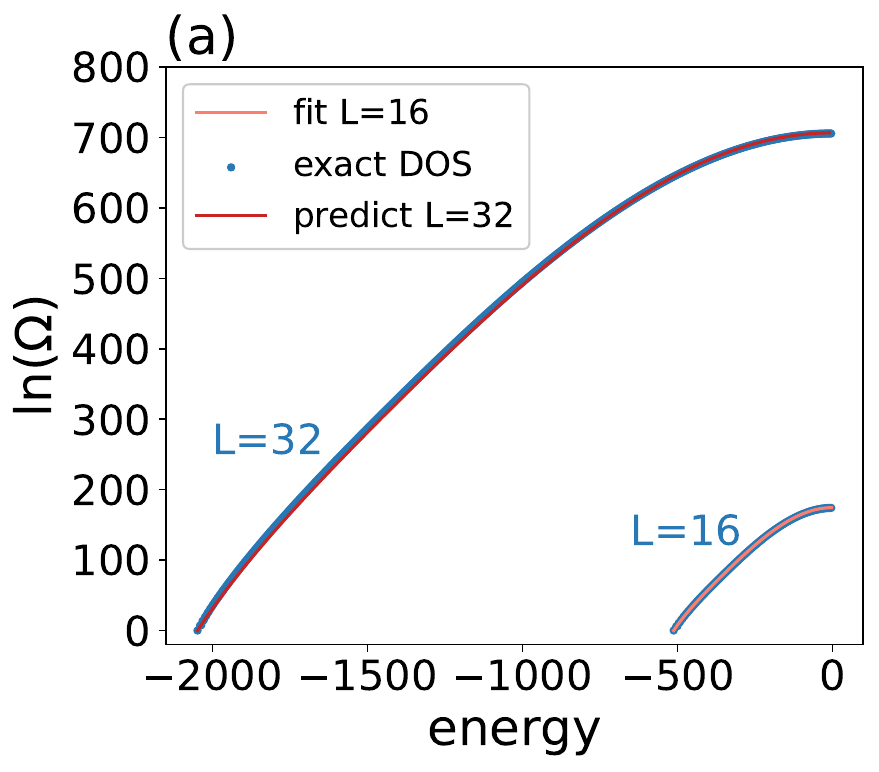}
\includegraphics[width=0.465\linewidth]{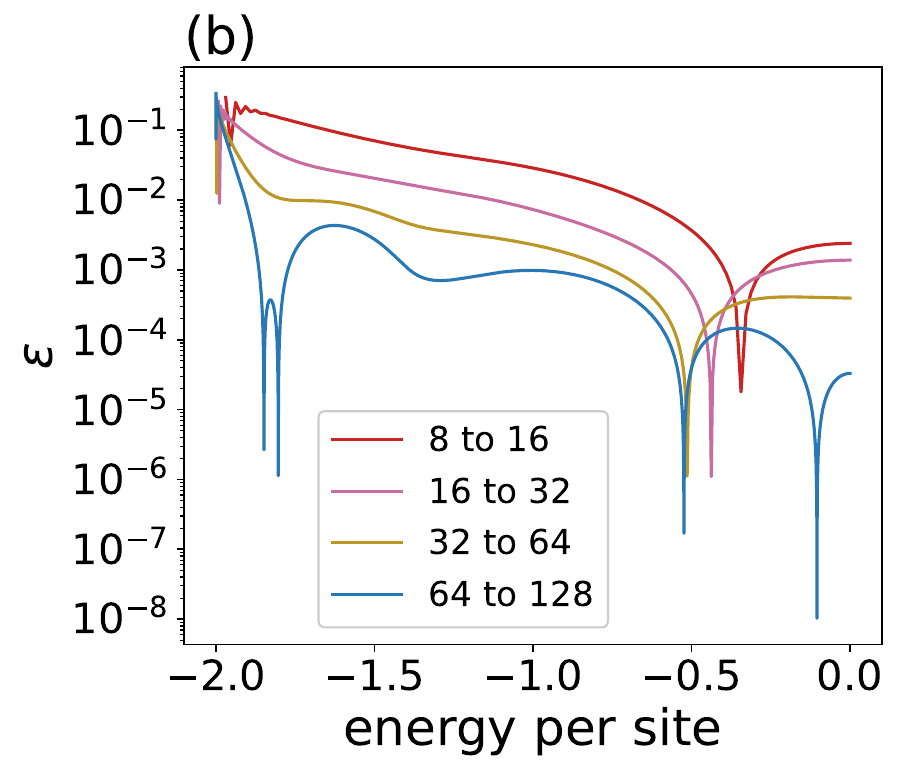}
\includegraphics[width=0.475\linewidth]{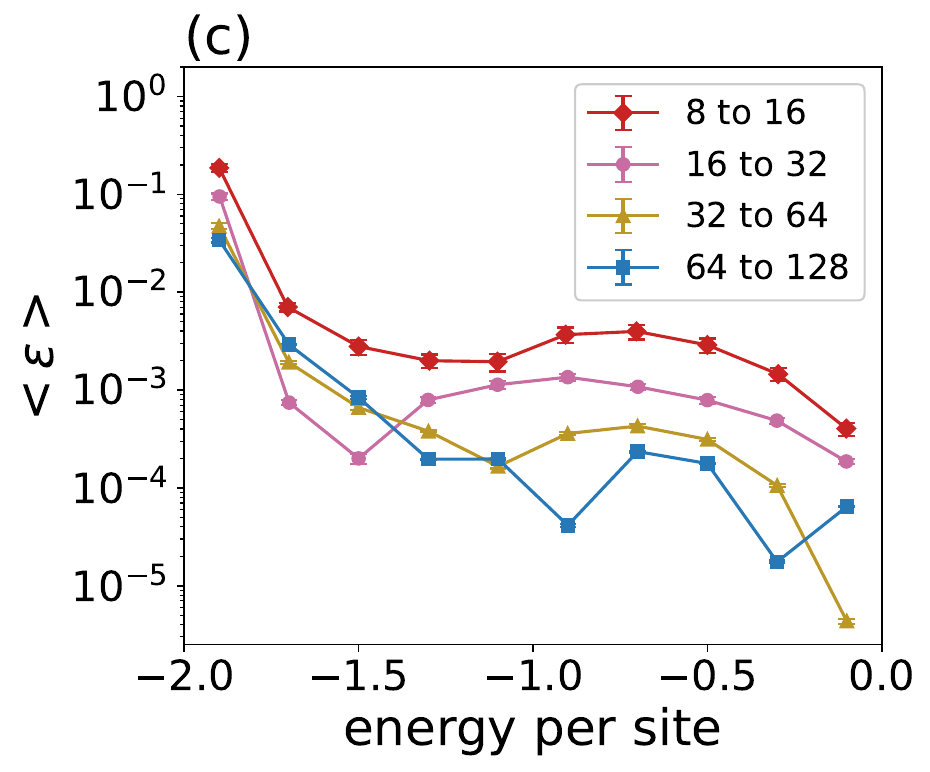}
\includegraphics[width=0.46\linewidth]{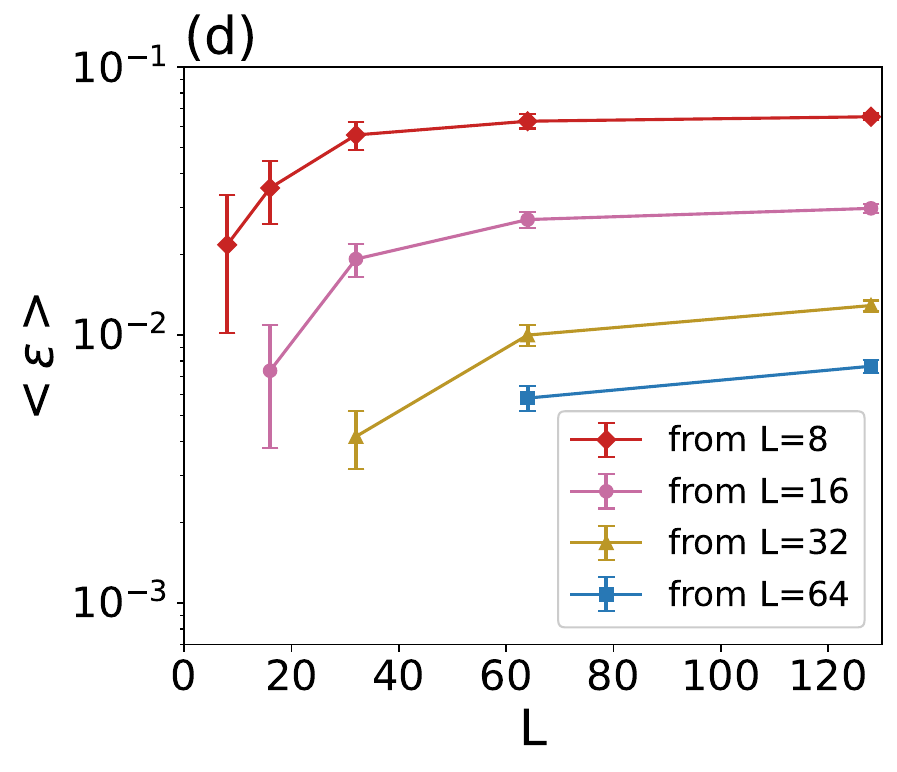}
\caption{(a) The figure shows the logarithmic exact density of states for the 2D Ising model at \(L=16\) and \(L=32\), along with polynomial fits to the logarithmic density of states at \(L=16\) and the logarithmic density of states at \(L=32\) estimated by density state curve mapping. The estimated density of states aligns closely with the exact density of states across most energy regions. (b) The figure shows the percentage error after doubling the density of states based on different system sizes, with large errors concentrated near the lowest energies. (c) This figure also shows the errors after doubling the density of states based on different system sizes, but before calculating the errors, the entire energy range is divided into 10 regions. A constant is added to the estimated density of states within each region to match its minimum value with the exact density of states, and then the average percentage error across the entire energy region is calculated. Compared to (b), the overall error is reduced by an order of magnitude. (d) The figure presents the average percentage error of the estimated density of states when the base system size is expanded to different system sizes. As the base system size increases, the average error of the estimated density of states is lower. }
\label{fig:fig6}
\end{figure}

The pink curve in Fig. \ref{fig:fig6}(a) shows the results of fitting the density of states for \(L=16\) using a 10th-order polynomial. Using a higher order polynomial does not sufficiently improve the result in the present case. For more complex systems, a higher order polynomial may be needed.
The red line in Fig. \ref{fig:fig6}(a) represents the predicted density of states for \(L=32\) using the polynomial fitted to \(L=16\). Within the energy range of \(E \in [-500, 0]\), the predicted density of states matches well with the analytical density of states \cite{beale1996exact}. However, as the energy decreases, the predicted density of states begins to deviate from the exact values, specifically, the predicted density of states is lower than the exact density of states.~\\

To quantify the deviation in predicted density of states, we examined the relationship between the percentage error of the predicted density of states and the energy, where the percentage error is defined as follows \cite{belardinelli2007fast},
\begin{equation}
\epsilon(E) = \frac{|\ln\lbrack\Omega_{pre}(E)\rbrack - \ln\lbrack\Omega_{exact}(E)\rbrack|}{\ln\lbrack\Omega_{exact}(E)\rbrack},
\end{equation}
As shown in Fig. \ref{fig:fig6}(b), except for some local dips, the percentage error in the predicted density of states in general increases with decreasing energy, regardless of the size of the base system. This observation aligns with the results seen in Fig. \ref{fig:fig6}(a) and can be attributed to the more complex variation trends in density of states at lower energies, which makes it challenging for the mapping method to accurately predict this region. With an increase in the size of the base system, the percentage error in predicted density of states shows a decreasing trend. This is because the larger the base system, the closer its density of states approaches the convergent value, as observed in Fig. \ref{fig:fig5}(a).~\\

The error in the predicted density of states, as shown in Fig. \ref{fig:fig6}(b), primarily originates from deviations at the lowest energies. Specifically, for the density of states curve of the given system size, the density of states in the low-energy region increases more rapidly with energy as compared to the high-energy region. When using polynomial fit for the entire density of states curve, the optimization process tends to increase the density of states values across the whole energy range to ensure a good fit in the low-energy region. This introduces deviations in the high-energy region, increasing the overall error. However, despite the overall increase in values, the local curve fitting in the high-energy region still performs well.
To mitigate the error induced by the deviation in the low energy region, we divide the energy range into 10 regions with intervals of 0.2. Within each region, we add a constant to the predicted density of states to match the minimum value of the predicted density of states with that of the exact density of states. We then compute the average percentage error within each region, as the results are presented in Fig. \ref{fig:fig6}(c). This error estimation helps us better assess the accuracy of the predicted density of states since, in Multi-cannocial Monte Carlo or MacroMC, we only need the density of states within a small energy range. From Fig. \ref{fig:fig6}(c), it is evident that the predicted density of states in the range \(E/N \in [-2, -1.8]\) has the highest percentage error, which quickly decreases to between \(10^{-3}\) and \(10^{-5}\), making the segmented errors approximately ten times lower than those shown in Fig. \ref{fig:fig6}(b). As the system size increases, the computational time required to achieve such high precision in estimated density of states through sampling methods would exponentially grow.
The method presented here only requires simple polynomial fitting and thus can substantially reduce the time consumed to estimate the density of states in large systems as compared to the sampling methods. 
~\\

In addition to doubling the system size, we also investigated the performance of our algorithm to estimate the density of states in system that are of multiples of the base system size. As shown in Fig. \ref{fig:fig6}(d), irrespective of the base system size, the average error in segmented energy increases with the targeted system size and eventually converging to a fixed value. Moreover, as the size of the base system increases, the error in the predicted density of states for larger systems converges faster and to a smaller error. We estimate that when the base system size reaches or exceeds 100, the error in the predicted density of states no longer increases significantly with the times of expansion, implying that high precision density of states for extremely large system sizes can be easily obtained.~\\

\begin{figure} [t!]
\includegraphics[width=0.46\linewidth]{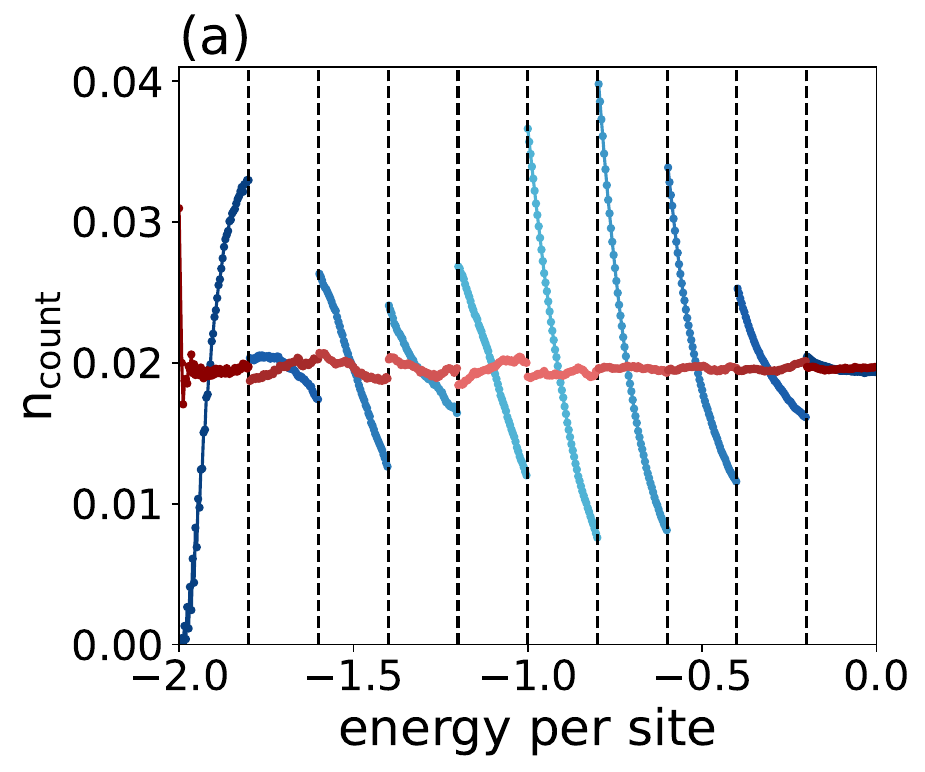}
\includegraphics[width=0.46\linewidth]{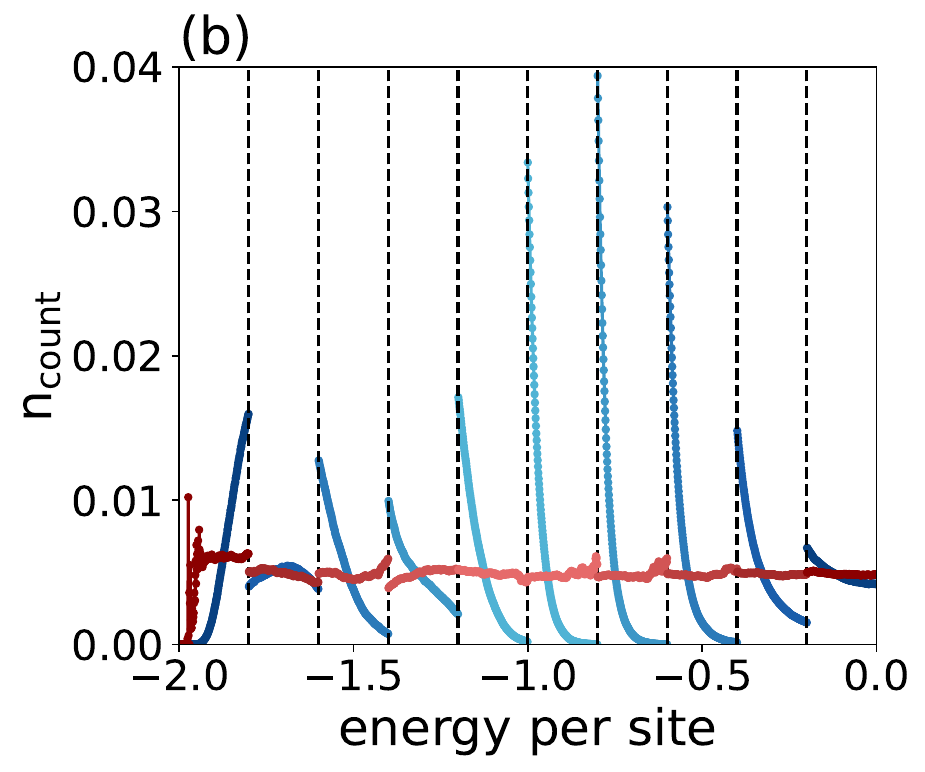}
\caption{Flat energy histogram sampling results for the 2D Ising model for (a) \(L=32\) and (b) \(L=64\), while the base system size is 16 for both of them. The blue and red curve represents the results from the density of states mapping before and after correction (see text for the details), respectively.}
\label{fig:fig7}
\end{figure}

Another method to demonstrate the accuracy of the predicted density of states is by using flat energy histogram sampling with the predicted density of states. As illustrated in Fig. \ref{fig:fig7}, we expanded the base system of \(L=16\) once to obtain the predicted density of states for \(L=32\) (Fig. \ref{fig:fig7}(a)) and twice to obtain the predicted density of states for \(L=64\) (Fig. \ref{fig:fig7}(b)). In each energy interval, we performed single-update Monte Carlo sampling with an acceptance rate \(R = \min\left(\frac{\Omega(S_{\text{before}})}{\Omega(S_{\text{after}})}, 1\right)\), where \(S_{\text{before}}\) and \(S_{\text{after}}\) represent the spin configurations before and after flipping, respectively, and \(\Omega(S_{\text{before}})\) and \(\Omega(S_{\text{after}})\) are the corresponding predicted density of states of these spin configurations. We take \(10^5\) sweeps of sampling in each energy interval and rescaled the results of the sampling,
\begin{equation}
n_{\text{count}}(E_{i}) = \frac{n_{\text{count}}(E_{i})}{\sum_{i = 0}^{M}{n_{\text{count}}(E_{i})}},
\label{eq10}
\end{equation}
where \(n_{\text{count}}(E_i)\) represents the number of times a macrostate \(E_i\) is sampled and \(M\) denotes the total number of macrostates \(E_i\) within the energy interval. The results, illustrated by the blue curve in Fig. \ref{fig:fig7}, indicate that despite the relatively larger errors in the predicted density of states at low energies, 
Monte Carlo sampling successfully covered the entire energy region without being trapped in any local energy region. This suggests that the predicted density of states in this region is sufficiently accurate to be used in the MacroMC simulations. In other regions with higher energy,  the resulting energy histograms are more flat. In the least flat region \(E \in [-0.8, -0.6]\), the maximum \(n_{\text{count}}(E = -0.8)\) is only about four times larger than the minimum \(n_{\text{count}}(E = -0.6)\). ~\\ 

Although the histograms in most energy intervals do not appear to be flat, the predicted density of states can still be considered as good. Figure \ref{fig:fig3}(c,d) can be an example to explain this. In Fig. \ref{fig:fig3}(c), the \(n_{\text{count}}(E \approx -1.55)\) is about 30 times greater than \(n_{\text{count}}(E \approx -0.9)\), which means the estimated density of states used in \ref{fig:fig3}(c) is much worse than that in Fig. \ref{fig:fig7}. However, we can still get a perfect Boltzmann distribution in \ref{fig:fig3}(d). The truly bad estimated density of states is the case where \(n_{\text{count}} = 0\) in a large energy range. This means the PMC sampling is trapped in a small energy region and cannot give us an unbiased result. More examples can be found in Appendix \ref{sec:Robustness_of_MacroMC}. Although the non-flat histograms are good enough for sampling, in the following, we will show how they can be corrected to a flat histogram to improve the performance.~\\

During the flat energy histogram sampling, an accurate density of states would result in a flat distribution of $n_{\text{count}}$. In fact, the non-uniform distribution can be used to further improve the estimated density of states. Specifically, we selected an energy interval in Fig. \ref{fig:fig7}, calculated the average value of $n_{\text{count}}$ within this interval, denoted it as \(\widetilde{n}_{\text{count}}\), and then compute the ratio of $n_{\text{count}}$ to \(\widetilde{n}_{\text{count}}\) for each energy \(E\). Taking the logarithm of this ratio gives us the correction factor \(C = \log(n_{\text{count}}/\widetilde{n}_{\text{count}})\) for the logarithmic density of states at each \(E\). The corrected logarithmic density of states is given by \(\log{(\Omega^{'})} = \log{(\Omega)} + C\). The red curve in Fig. \ref{fig:fig7} shows the flat energy histogram sampling using the corrected density of states, where it is evident that the distribution of $n_{\text{count}}$ is much more uniform across the entire energy range.~\\

\section{Conclusion}
\label{sec:conculsion}
In this paper, we propose a Monte Carlo method named MacroMC, that aimed at improving sampling efficiency and accuracy in condensed matter systems. First, we perform flat histogram sampling from PMC using the density of states to obtain proposed spin configurations. These spin configurations are then randomly selected in MacroMC. Next, we combine the density of states with Boltzmann weights to determine acceptance or rejection. Introducing the density of states significantly increases the probability of the Markov chain jumping to states with larger energy differences in a single step. This effectively overcomes the problem of getting trapped in local minima within complex energy landscapes and reduces autocorrelation between samples. Our results show that the autocorrelation of MacroMC in the 2D Ising model and the 2D 10-state Potts model is significantly smaller than that of the cluster-flip Monte Carlo. The autocorrelation in the 2D FF XY model and the 2D EA model is also significantly small. The algorithm has potential extensions to other lattice structures and higher-dimensional models. We also estimate the BKT transition temperature of the FF XY model and find that our result is slightly higher than previous results.~\\ 

In addition, we propose a mapping method based on the density of states of small systems. This method leverages the advantage of easily obtaining high-precision density of states in small systems. By analyzing the trend of changes in the density of states curve of small systems, we can estimate the density of states of larger systems, and segmental correction of the density of states can further improve the prediction accuracy. Our results show that the density of states for the \(L=16\) 2D Ising model can be mapped to the valuable density of states for \(L=32\) and \(L=64\). This mapping method avoids estimating the density of states in large systems through sampling, thus significantly saving computational resources.~\\

The mapping method provides us with the density of states for larger systems, allowing us to apply MacroMC to larger system sizes. The combination of these two methods enables us to explore the entire configuration space more efficiently and obtain more accurate estimates of physical quantities.~\\

For future work, one may apply the density of states mapping to quantum models such as the Hubbard model. By employing Monte Carlo algorithm for quantum system such as world-line quantum Monte Carlo and Determinant quantum Monte Carlo, we can obtain high-precision density of states for small system sizes where the sign problem is not severe \cite{batrouni1992world, santos2003introduction,assaad2008world}. By applying the proposed mapping method, we can then obtain the density of states for larger systems and further study the phase transitions of the system. Additionally, one can explore how to combine Macrostate Monte Carlo with quantum Monte Carlo to improve sampling efficiency in situations where the sign problem is severe.


\begin{acknowledgments}
We acknowledge financial support from Research Grants Council of Hong Kong (Grant No. CityU 11318722), National Natural Science Foundation of China (Grant No. 12005179, 12204130),
City University of Hong Kong (Grant No. 9610438, 7005610), Harbin Institute of Technology Shenzhen (Grant No. X20220001), Shenzhen Start-Up Research Funds~(No.~HA11409065) and Shenzhen Key Laboratory of Advanced Functional Carbon Materials Research and Comprehensive Application~(No.~ZDSYS20220527171407017).
\end{acknowledgments}

\nocite{*}

%

\appendix
\section{Detailed balance of proposal Monte Carlo}
\label{sec:Detailed_balance_of_proposal_monte carlo}

Consider two states in a Markov chain, \(s_{1}\) and \(s_{2}\), which belong to the macrostates with energies \(E_{1}\) and \(E_{2}\), and densities of states \(\Omega_{1}\) and \(\Omega_{2}\), respectively. To obtain a flat energy histogram, each macrostate should have the same occurrence probability of \(P_{E_{i}}=1/M\), where \(M\) is the total number of macrostates in the system. Each microstate in one macrostate has the same occurrence probability \(P_{s_{i}}=1/\Omega_{i}\). The occurrence probabilities of \(s_{1}\) and \(s_{2}\) among all macrostates are the product of \(P_{E_{i}}\) and \(P_{s_{i}}\), which are \(1/(\Omega_{1}M)\) and \(1/(\Omega_{2}M)\), respectively. The transition probabilities from state 1 to state 2 is \(\pi_{12}\) and that for the reverse process is \(\pi_{21}\). In PMC, \(\pi_{12}\) is reformulated as the product of the generation probability of \(s_{2}\) from \(s_{1}\), denoted as \(T_{12}\), and the acceptance probability of \(s_{2}\) given \(s_{1}\), denoted as \(A_{12}\), this yields \(\pi_{12} = T_{12}A_{12}\). To maintain a detailed balance, the PMC model must satisfy the following conditions:
\begin{equation}
P_{1}\pi_{12} = P_{1}T_{12}A_{12} = P_{2}\pi_{21} = P_{2}T_{21}A_{21},
\label{A1}
\end{equation}
and the acceptance rate \(R_{12}\) of PMC reads,
\begin{equation}
R_{12} = \frac{A_{12}}{A_{21}} = \frac{P_{2}T_{21}}{P_{1}T_{12}}.
\label{A2}
\end{equation}

The update mechanism of PMC is similar to that of MCMC - a random selection of a site and an attempt to flip its spin. If \(s_{1}\) and \(s_{2}\) differ by the spin at only one site, then \(T_{12}\) is the probability of randomly selecting this site, which is \(1/L^{2}\), where \(L\) is the linear system size. Similarly, \(T_{21}\) is also \(1/L^2\). When \(s_{1}\) and \(s_{2}\) differ by the spins at two or more sites, \(s_{1}\) cannot be transformed into \(s_{2}\) by a single update, hence \(T_{12}\) equals 0, and similarly, \(T_{21}\) equals 0. In both cases, \(T_{12} = T_{21}\) holds, making the acceptance rate \(R_{12}\) dependent on the occurrence probability of each state in the Markov chain,
\begin{equation}
R_{12} = \frac{A_{12}}{A_{21}} = \frac{P_{2}}{P_{1}} = \frac{\Omega_{1}}{\Omega_{2}}.
\label{A3}
\end{equation}

\section{Sampling with the estimated density of states}
\label{appB}

\begin{figure} [t!]
\centering
\includegraphics[width=0.8\linewidth]{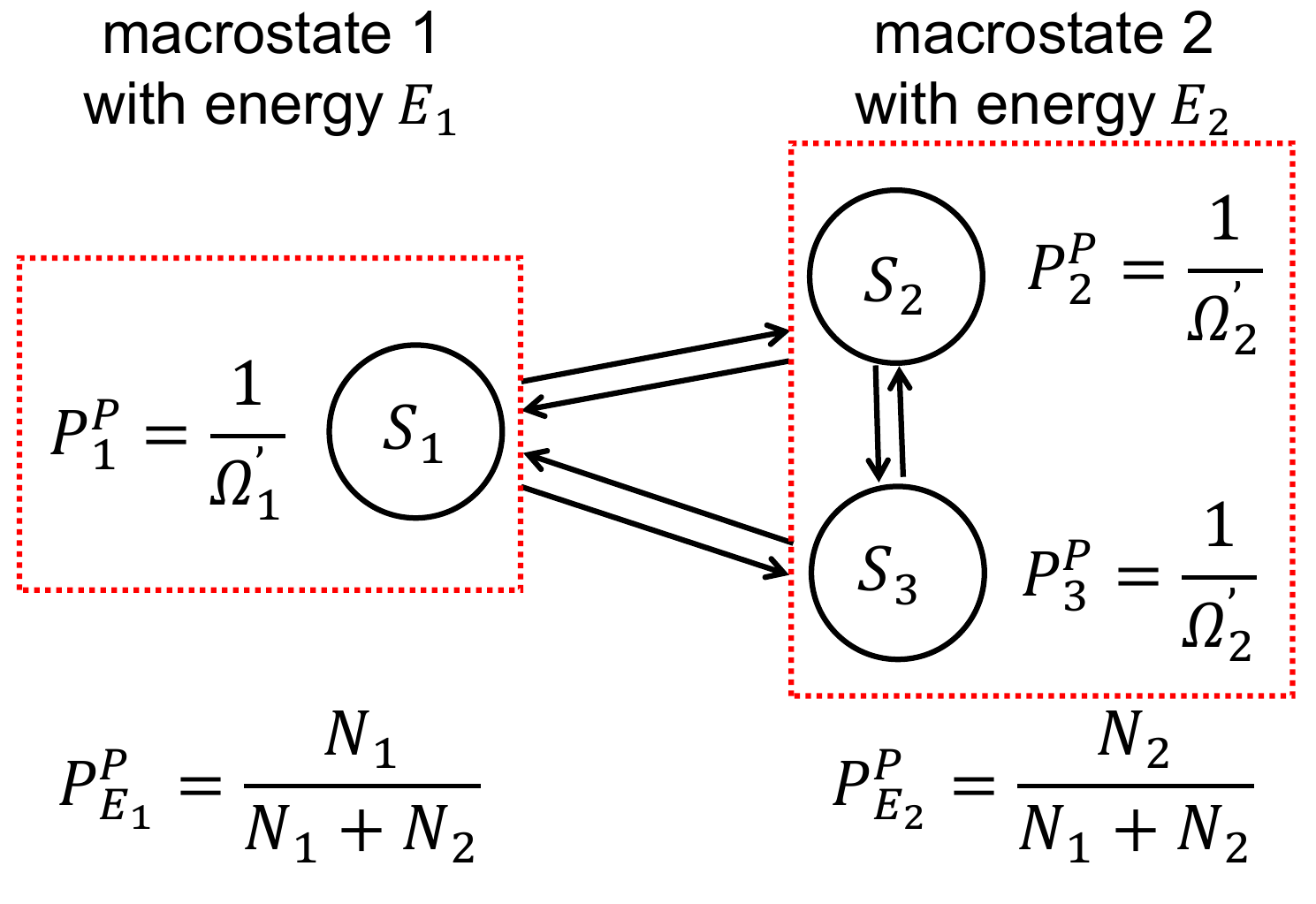}
\caption{The figure shows an artificial system being sampled by PMC, where circles represent three microstates \(s_{1}\), \(s_{2}\), and \(s_{3}\) of the system, and red dashed rectangles denote two macrostates with energy \(E_{1}\) and \(E_{2}\). The occurrence probability of each microstate is \(P^P_{1}\), \(P^P_{2}\), and \(P^P_{3}\), and the occurrence probability of each macrostate is \(P^P_{E_{1}}\) and \(P^P_{E_{2}}\).}
\label{fig:fig8}
\end{figure}

\label{sec:Sampling_with_non-exact_density_of_state}
When using the exact density of states for PMC sampling, it can produce a totally flat energy histogram. However, with an estimated density of states, the resulting energy histogram from PMC sampling will depend on the difference between the estimated and the exact density of states. We will explain this in detail in the following. Note that we will adopted the notation \(P^{P}\) and \(P^{M}\) for the occurrence probability of a state in PMC and MacroMC sampling, respectively.~\\

Let us illustrate the PMC process with an artifical system shown in Fig. \ref{fig:fig8}. Assume there are three microstates in the Markov chain, \(s_{1}\), \(s_{2}\), and \(s_{3}\), with \(s_{1}\) belonging to the macrostate at energy \(E_{1}\), \(s_{2}\) and \(s_{3}\) belonging to the macrostate at energy \(E_{2}\). For PMC sampling, the occurrence probabilities of these microstates are \(P_{1}^{P}=1/\Omega_{1}'\) and \(P_{2}^{P}=P_{3}^{P}=1/\Omega_{2}'\), where $\Omega_{1}'$ and $\Omega_{2}'$ is the estimated density of states for macrostates with energy $E_{1}$ and $E_{2}$, respectively. The probability of each macrostate is \(P_{E_{1}}^{P}=N_{1}/(N_{1}+N_{2})\) and \(P_{E_{2}}^{P}=N_{2}/(N_{1}+N_{2})\), where $N_i$ is the the number of samples having energy $E_i$ in PMC. If we know the exact values of the density of states, in this case, we have \(\Omega_{1}'=1\), \(\Omega_{2}'=2\), and $N_{1}=N_{2}$, therefore \(P_{E_{1}}^{P}=P_{E_{2}}^{P}=1/2\). However, in reality, the exact values of the density of states are unknown.~\\

At detailed balance of the PMC sampling, \(T_{ij} = T_{ji}\), leading to the following,
\begin{equation}
\begin{aligned}
& P_{1}^{P}\pi_{12} = P_{1}^{P}A_{12} = P_{2}^{P}\pi_{21} = P_{2}^{P}A_{21}, \\
& P_{1}^{P}\pi_{13} = P_{1}^{P}A_{13} = P_{3}^{P}\pi_{31} = P_{3}^{P}A_{31}. \\
\end{aligned}
\label{B1}
\end{equation}
The probability of each macrostate is the sum of the probabilities of its microstates, i.e.
\begin{equation}
\begin{aligned}
& P_{E_{1}}^{P} = \Omega_{1}P_{1}^{P}, \\
& P_{E_{2}}^{P} = P_{2}^{P} + P_{3}^{P} = \Omega_{2}P_{2}^{P}, \\
\end{aligned}
\label{B2}
\end{equation}
here we use the condition that $P_{2}^{P}=P_{3}^{P}$. The ratio of the probabilities to find the two macrostates can be derived as follow,
\begin{equation}
\frac{P_{E_{1}}^{P}}{P_{E_{2}}^{P}} = \frac{\Omega_{1}P_{1}^{P}}{\Omega_{2}P_{2}^{P}} = \frac{\Omega_{1}P_{2}^{P}\frac{A_{21}}{A_{12}}}{\Omega_{2}P_{2}^{P}} = \frac{\Omega_{1}A_{21}}{\Omega_{2}A_{12}}=\frac{\Omega_{1}\Omega_{2}^{'}}{\Omega_{2}\Omega_{1}^{'}}.
\label{B3}
\end{equation}
Note that we have used $P_{1}^{P}=P_{2}^{P}A_{12}/A_{21}$ from Eq. \ref{B1}.~\\

When sampling from a non-flat energy histogram using MacroMC, the ratio of the occurrence probabilities for any two macrostates in the Markov chain is given as follow,
\begin{equation}
\frac{P_{E_{i}}^{M}}{P_{E_{j}}^{M}} = \frac{T_{ji}A_{ji}}{T_{ij}A_{ij}} = \frac{T_{ji}\Omega_{i}^{'}W_{i}}{T_{ij}\Omega_{j}^{'}W_{j}},
\label{B5}
\end{equation}
where \(W_{i}\) represents the Boltzmann weight. As mentioned in the main text, \(T_{ij}\) in MacroMC is the probability of selecting a spin configuration with energy \(E_j\) from the spin configuration pool obtained through PMC sampling, in other words, \(T_{ij}\) is the occurrence probability of a macrostate with energy $E_{j}$ in PMC, which is $P_{E_{j}}^{P}$. Similar argument applies for \(T_{ji}\), leading to the ratio
\begin{equation}
\frac{T_{ji}}{T_{ij}} = \frac{P_{E_{i}}^{P}}{P_{E_{j}}^{P}} = \frac{\Omega_{i}\Omega_{j}^{'}}{\Omega_{j}\Omega_{i}^{'}}.
\label{B6}
\end{equation}
Consequently, the occurrence ratio of any two macrostates in MacroMC is
\begin{equation}
\frac{P_{E_{i}}^{M}}{P_{E_{j}}^{M}} = \frac{\Omega_{i}\Omega_{j}^{'}}{\Omega_{j}\Omega_{i}^{'}} \frac{\Omega_{i}^{'}W_{i}}{\Omega_{j}^{'}W_{j}} = \frac{\Omega_{i}W_{i}}{\Omega_{j}W_{j}}.
\label{B7}
\end{equation}
The final result allows us to measure the physical quantities that satisfy thermodynamic equilibrium even with an inaccurate density of states.~\\

\section{Robustness of MacroMC}
\label{sec:Robustness_of_MacroMC}

\begin{figure*}[t!]
\centering
\includegraphics[width=0.32\linewidth]{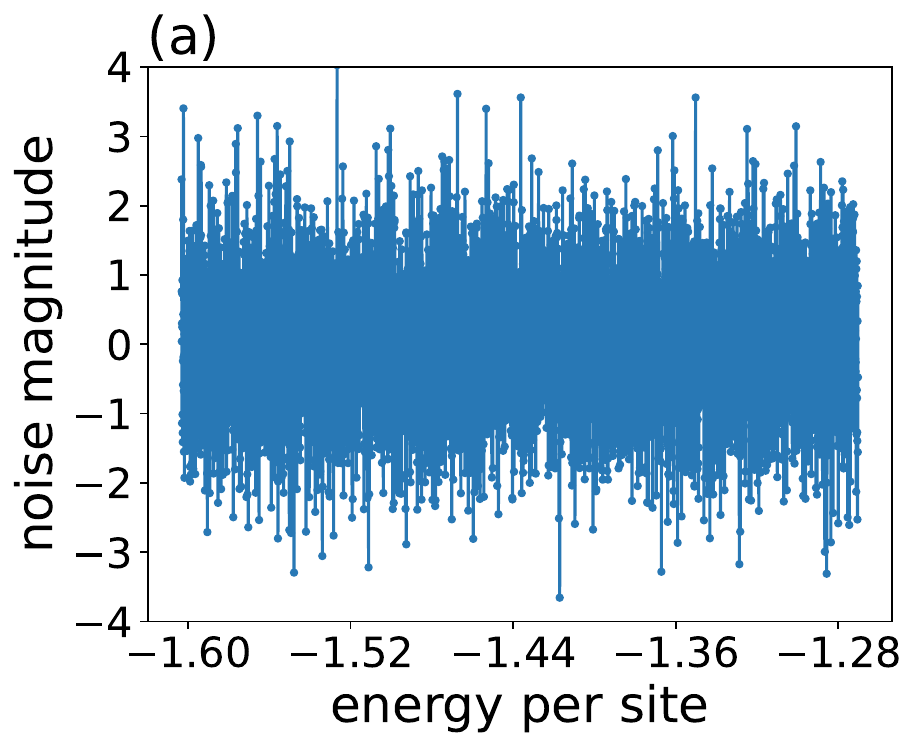}\hfill
\includegraphics[width=0.335\linewidth]{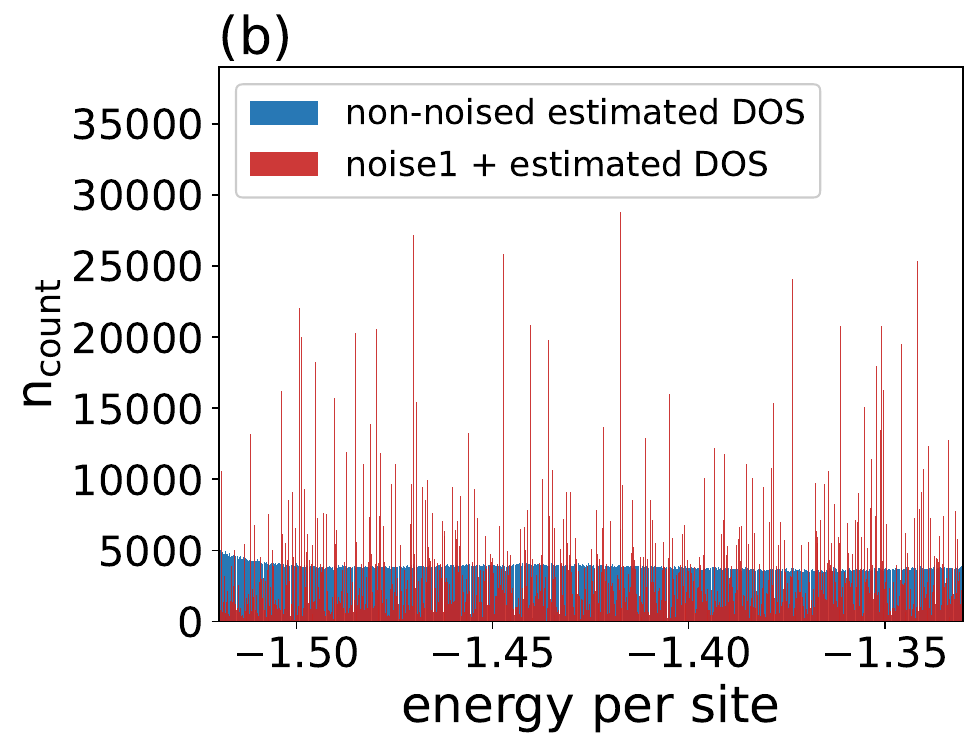}\hfill
\includegraphics[width=0.325\linewidth]{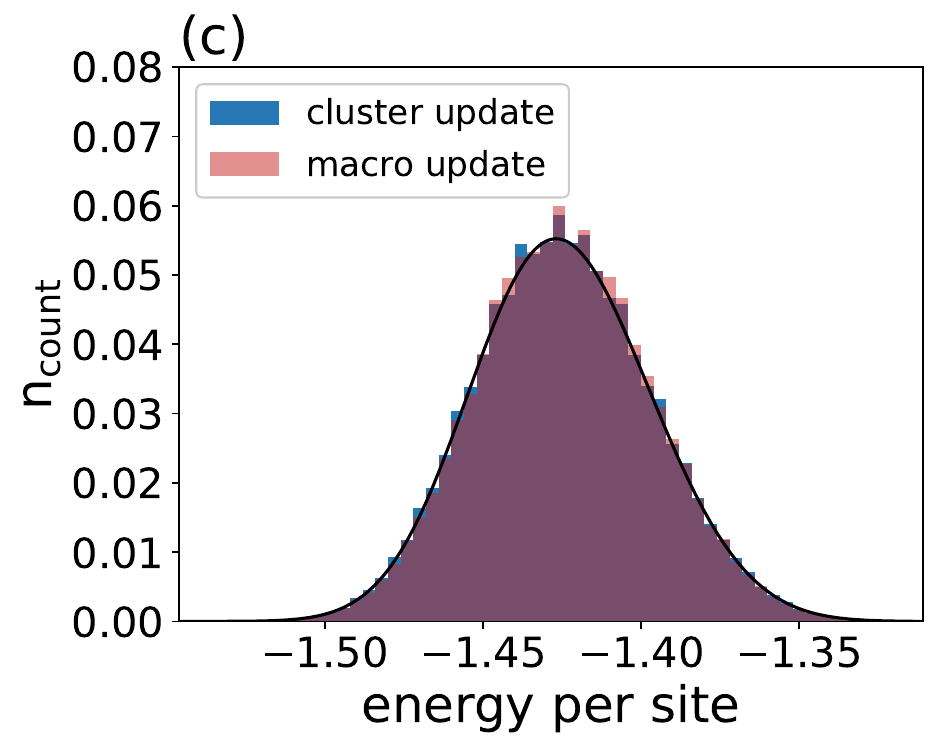}
\bigskip 
\includegraphics[width=0.32\linewidth]{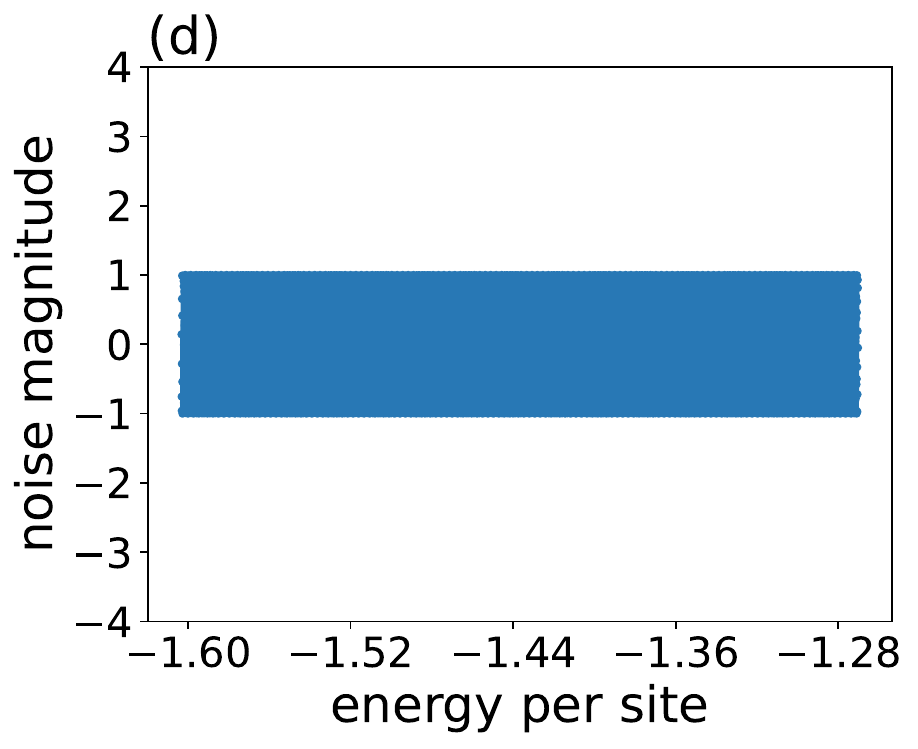}\hfill
\includegraphics[width=0.335\linewidth]{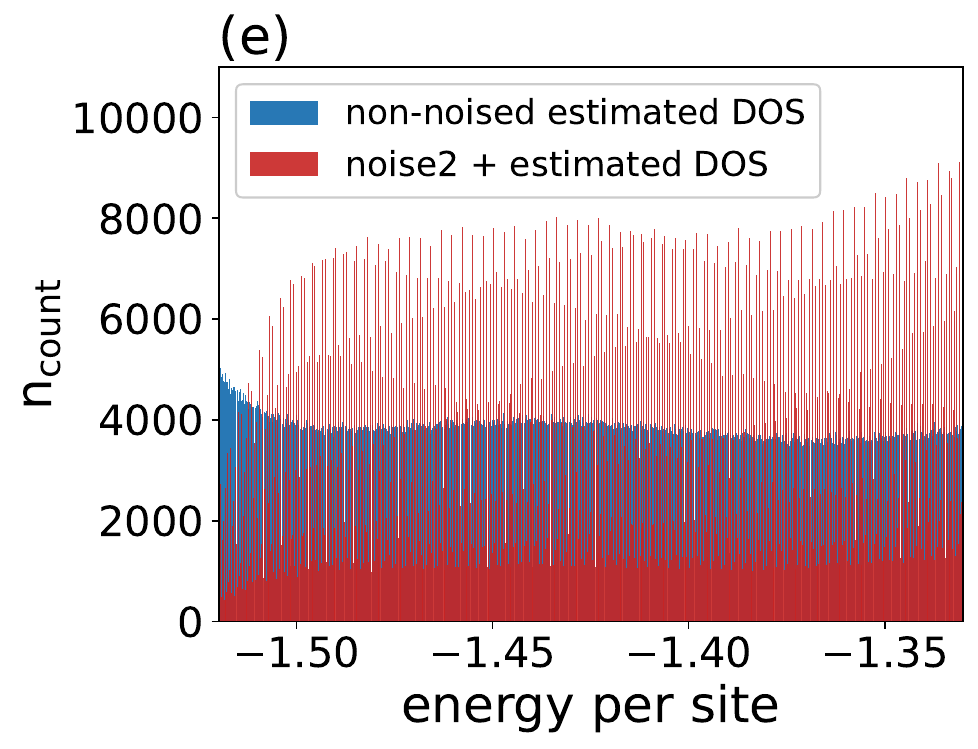}\hfill
\includegraphics[width=0.325\linewidth]{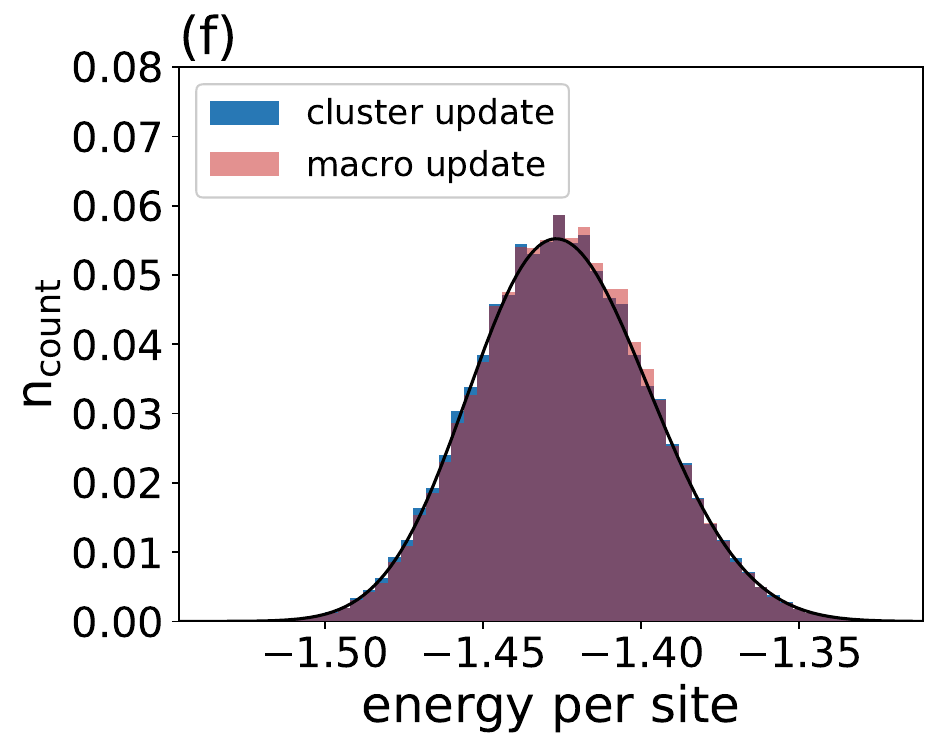}
\includegraphics[width=0.32\linewidth]{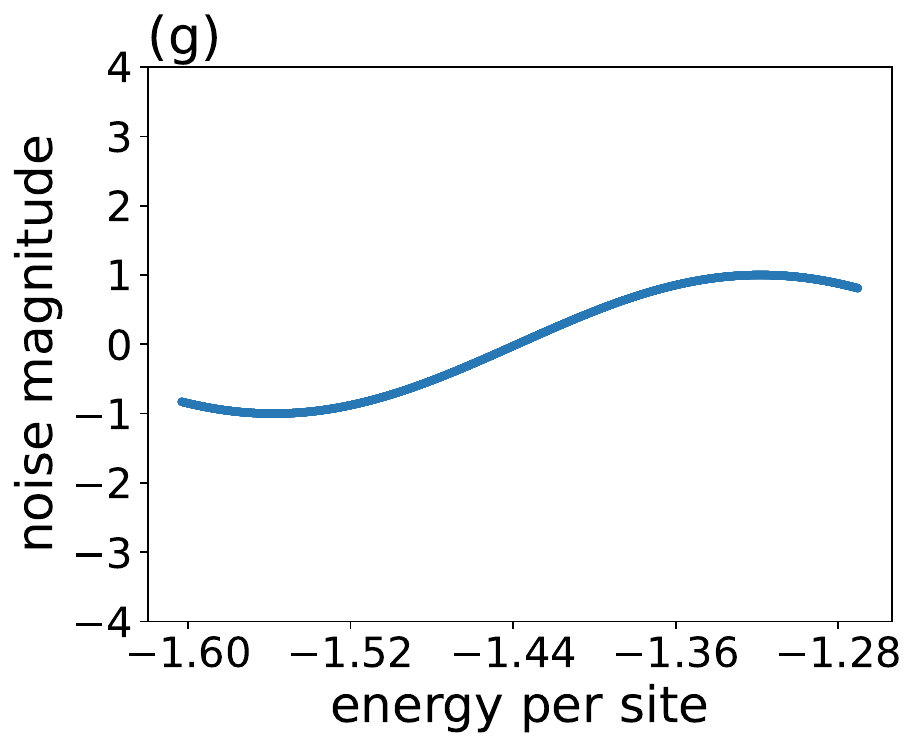}\hfill
\includegraphics[width=0.335\linewidth]{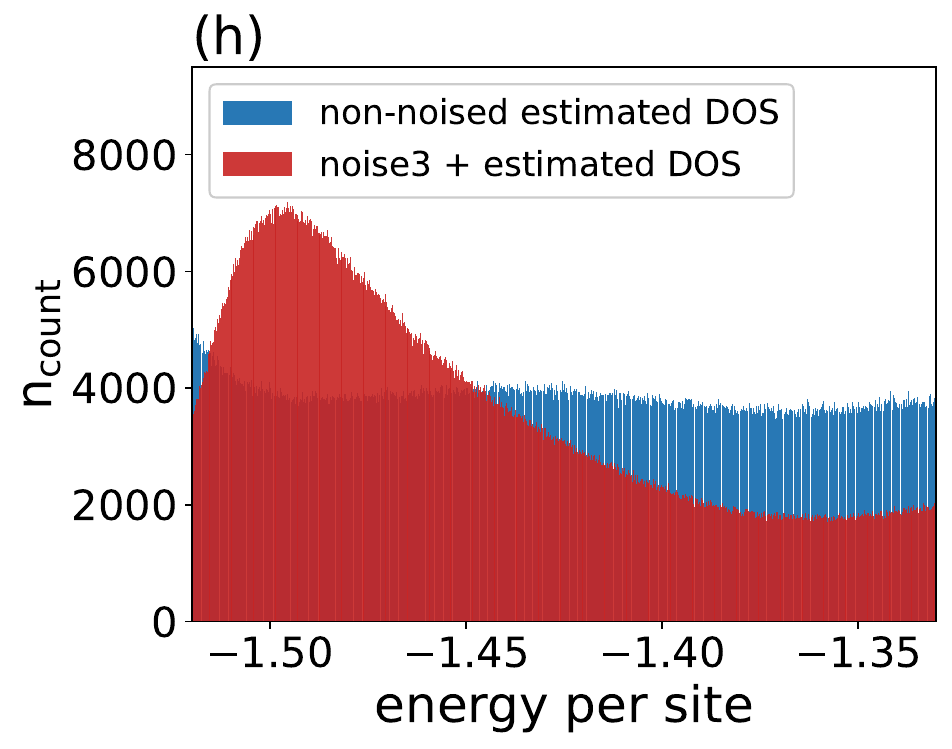}\hfill
\includegraphics[width=0.325\linewidth]{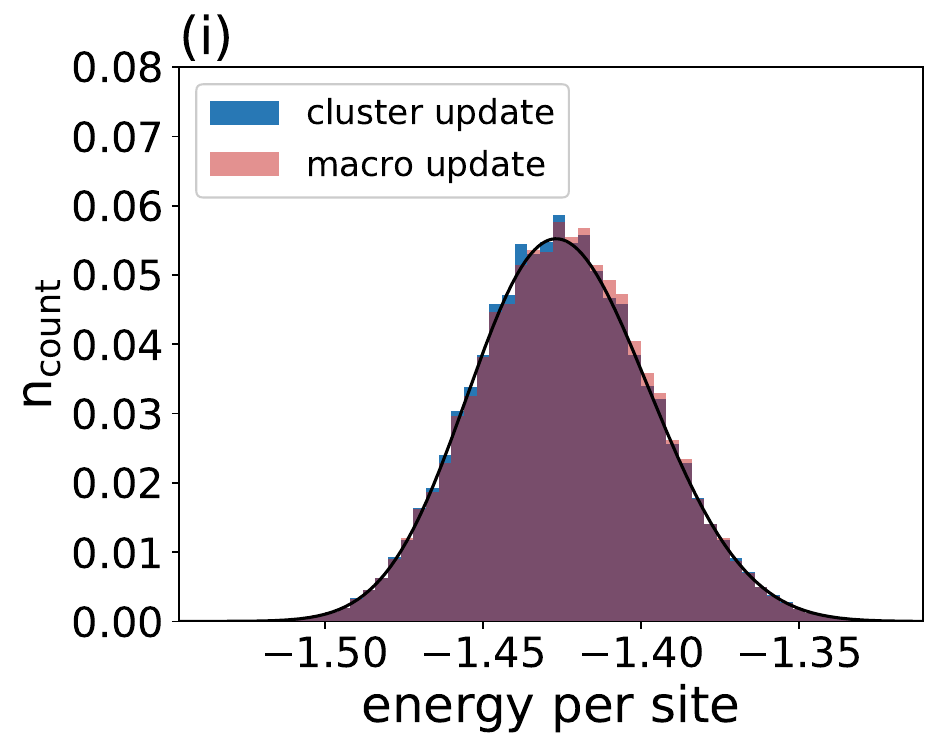}
\caption{(a), (d), and (g) show three types of noise added on the estimated density of states. (b), (e), and (h) show the sampling results of PMC for the 2D Ising model using both noise-free and noisy estimated density of states. (c), (f), and (i) compare the Boltzmann sampling results of the 2D Ising model using cluster update and MacroMC with the noisy estimated density of states.}
\label{fig:fig9}
\end{figure*}

We demonstrated the robustness of MacroMC on a 2D Ising model with \(L=128\) under the transition temperature at \(T = 2.267\). Using the scheme discussed in Sec. \ref{sec:density_state_curve_mapping}, we can easily obtain a relatively accurate estimated density of states. When using the estimated density of states for sampling with PMC, it produces a nearly flat energy histogram, as illustrated by the blue histogram in Fig. \ref{fig:fig9}(b, d, h). When strong local Gaussian fluctuations  \({\text{noise1}} = \mathcal{N}(0,1)\) or sinusoidal \({\text{noise2}} = \sin(E)\) are added to the estimated density of states, the energy histograms sampled through PMC exhibit many extreme values, as shown in Fig. \ref{fig:fig9}(b,e). However, even when sampling from these highly fluctuating energy histograms, MacroMC can achieve results comparable to those obtained with cluster updates. When noise with a global trend \({\text{noise3}} = \sin(0.0008E)\) is added to the estimated density of states, the energy histogram sampled through PMC also adopts this global trend. Similarly, this global trend does not significantly impact the sampling accuracy of MacroMC.

\end{document}